\documentclass[%
 aip,
amsmath,amssymb,
 reprint,%
]{revtex4-1}

\usepackage{graphicx,color}
\usepackage{dcolumn}
\usepackage{bm}

\usepackage[utf8]{inputenc}
\usepackage[T1]{fontenc}
\usepackage{mathptmx}
\usepackage{etoolbox}
\usepackage{soul}

\usepackage{mathtools}
\newcommand\myeq{\stackrel{\mathclap{\varphi = \pi/2}}{=}}
\newcommand\myeqb{\stackrel{\mathclap{?}}{=}}

\makeatletter
\def\@email#1#2{%
 \endgroup
 \patchcmd{\titleblock@produce}
  {\frontmatter@RRAPformat}
  {\frontmatter@RRAPformat{\produce@RRAP{*#1\href{mailto:#2}{#2}}}\frontmatter@RRAPformat}
  {}{}
}%
\makeatother

\begin{document}

\preprint{AIP/123-QED}

\title{Optimal asymptotic precision bounds for nonlinear quantum metrology under collective dephasing} 

\bigskip

\author{Francisco Riberi}
 \affiliation{\mbox{Department of Physics and Astronomy, Dartmouth College, 6127 Wilder Laboratory, Hanover, New Hampshire 03755, USA}} 

\author{Lorenza Viola}
\affiliation{\mbox{Department of Physics and Astronomy, Dartmouth College, 6127 Wilder Laboratory, Hanover, New Hampshire 03755, USA}} 

\date{\today}

\begin{abstract}
Interactions among sensors can provide, in addition to entanglement, an important resource for boosting the precision in quantum estimation protocols. Dephasing noise, however, remains a leading source of decoherence in state-of-the-art quantum sensing platforms. We analyze the impact of classical {\em collective dephasing with arbitrary temporal correlations} on the performance of generalized Ramsey interferometry protocols with \emph{quadratic} encoding of a target frequency parameter. The optimal asymptotic precision bounds are derived for both product coherent spin states and for a class of experimentally relevant entangled spin-squeezed states of $N$ qubit sensors. While, as in linear metrology, entanglement offers no advantage if the noise is Markovian, a precision scaling of $N^{-1}$ is reachable with classical input states in the quadratic setting, which is improved to $N^{-5/4}$ when temporal correlations are present and the Zeno regime is accessible. The use of nonclassical spin-squeezed states and a nonlinear readout further allows for an $N^{-3/2}$ precision scaling, which we prove is asymptotically optimal. We also show how to counter {\em noise-induced bias} by introducing a simple ratio estimator which relies on detecting two suitable system observables, and show that it remains asymptotically unbiased in the presence of dephasing, without detriment to the achievable precision.
\end{abstract}

\maketitle

\section{Introduction}

Quantum sensing and metrology allow for advantage in estimation precision over classical approaches to be achieved by harnessing nonclassical resources, which may range from entanglement and squeezing to many-body interactions  \cite{Degen,SmerziRMP,RDDreview,BraunRMP}, or may even be tied to more exotic effects stemming from quantum criticality,  topology, or non-Hermiticity \cite{Zanardi,Budich,Clerk}. While quantum metrology has already found application in fields as diverse as atomic spectroscopy \cite{Wineland1,Wineland2}, magnetometry and thermometry \cite{Didi,Jones,LukinT}, time-keeping with atomic clocks \cite{clocks,Ye,Colombo} and gravitational-wave astronomy \cite{Gravity}, additional connections and implications for quantum science continue to emerge as experimental capabilities improve \cite{Essay}. From a theoretical standpoint, making the quantum advantage expected for these applications as substantial and robust as possible calls for a continued effort in quantifying and mitigating the impact of noise, that is inevitably present in reality.

In the most common single-parameter \emph{linear} estimation setting, the parameter of interest is coupled to an observable of $N$ qubit sensors which are \emph{non-interacting} following initialization. In the ideal noiseless limit, the optimal scaling of precision with the sensors' number that is achievable using product (classical) states is set by the $N^{-1/2}$ \emph{standard quantum limit} (SQL), which can be reached by preparing a coherent spin state (CSS). The ultimate precision bound is instead provided by a $N^{-1}$ scaling, also referred to as the \emph{Heisenberg limit} (HL) and attainable by employing an $N$-partite entangled Greenberger-Horne-Zeilinger (GHZ) state \cite{Sisi}. Noise is well-known to reduce or altogether preclude the precision advantage these bounds predict \cite{Escher,HaaseNJP}. In practice, one of the most pervasive sources of noise arises when the coupling to the noisy environment is mediated by an operator that commutes with the Hamiltonian that encodes the signal, leading to pure dephasing. In particular, perfectly spatially correlated, \emph{collective dephasing} arises when all the sensors couple identically to the noise source. This kind of noise is both practically relevant as it is particularly adversarial. Experimentally, it is a leading source of decoherence in the main platforms currently used for entanglement-assisted sensing, including Bose-Einstein condensates (BECs) \cite{Khor2009,BECCollDeph}, trapped-ion lattices and ensembles of cold atoms \cite{Carnio2015,Dorner2012,BEC}. Theoretically, is it known that if no temporal correlations are present and the dynamics are Markovian, the best possible precision has no SQL scaling, unlike spatially uncorrelated noise \cite{Huelga1997}, and need not even approach zero for large $N$ \cite{Dorner2012}; even in the more favorable scenario where noise is temporally correlated and the dynamics are non-Markovian \cite{Chin2012}, one can show that no scaling better than $N^{-1/2}$ is permitted \cite{third}. Likewise, due to its fully ``parallel'' nature, collective dephasing is not amenable to mitigation strategies using quantum error correction \cite{Layden2} or dynamical decoupling \cite{Sekatski,DurDD}, nor does it benefit from optimization over the sensors' initial state \cite{Jeske2014} or from randomizing or tailoring the sensors' geometry \cite{FUR, Bias}.

In addition to the above-assumed linear estimation framework, a different strategy for improving measurement precision consists on allowing for the sensors to interact {\em during} the signal imprint, such that nonclassical correlations may be generated after state preparation, possibly on top of pre-existing entanglement. Such a \emph{nonlinear} metrology strategy was introduced in Ref.\,\onlinecite{Luis2004} and extensively studied in later works \cite{DattaPRL,DattaPRA,Choi,Rey2007}. Notably, for a closed system where the sensing Hamiltonian allows for up to $k$th-body interactions ($k$-th order non-linearities, with $k>1$), the optimal ``classical'' precision scaling achievable with product states is a ``generalized SQL'' of the form $N^{-k+1/2}$, which is super-Heisenberg if compared to the best possible linear bound, $k=1$. If entangled initial states are used, the asymptotic precision is further predicted to attain a ``generalized HL'' of the form $N^{-k}$. 

While the implementation of highly nonlinear protocols is practically challenging, nonlinear metrology through quadratic encoding ($k=2$) has been realized experimentally in a light-matter interface \cite{Napolitano2011,Napolitano2014}, as well as in a nuclear magnetic resonance system \cite{Du2018} for moderately large $N$. In addition, BECs \cite{BoixoNonLinearBEC,Choi} and cold-atom ensembles \cite{Chase} have been extensively studied as candidates for obtaining the above advantageous scalings. A factor hindering a broader applicability of nonlinear metrology has been the fact that most of the existing schemes only work for the measurement of intrinsic properties of the sensing system itself (such as spin alignment or magnetization \cite{Napolitano2011,Napolitano2014}). However, major steps forward have recently been made to overcome these challenges. On the one hand, advances in the manipulation of cold-atom and molecule ensembles \cite{Browaeys2020, deMille2024} offer a greater degree of control over the sensors' correlations, which can be harnessed for engineering nonlinearities; on the other hand, theoretical proposals for measuring \emph{external fields} by means of a nonlinear setup have been put forth \cite{Generating,Deng2021}, 

In all of the above nonlinear settings, collective dephasing remains a leading source of decoherence. Yet, its impact on asymptotic precision scaling has received little theoretical attention thus far, and only in the Markovian regime to the best of our knowledge. Specifically, unlike for linear metrology, it was found that fully uncorrelated dephasing does not degrade the asymptotic precision scaling \cite{DelCampo,DattaPRA,Rey2007}; similarly, the scaling of precision was predicted to be unchanged under collective collisional decoherence in a \emph{phase} (not frequency) estimation setting, when working with an optimal coherent state \cite{BECCollDeph}. Even less attention has been payed to the fact that, as in the linear case \cite{Bias}, dephasing introduces a systematic bias in the standard noiseless estimator that is employed in the context of {\em local} estimation via the method of moments \cite{SmerziRMP}. In order to perform a successful sensing protocol, such a bias needs to be corrected or otherwise accounted for. Thus, it is important to derive a strategy capable of performing \emph{accurate} frequency estimation when interactions among sensors are allowed.
 
The present work aims to fill these gaps. We study the performance of a class of Gaussian (properly squeezed) one-axis-twisted states (OATS), which fall in a ``low-excitation regime'' where a Holstein-Primakoff mapping to a continuous bosonic degree of freedom may be faithfully carried out. In the context of linear metrology, this class of states has been shown to be more robust against preparation errors than other entangled initial conditions \cite{SchulteEchoes,OATS2023}, thereby making it more experimentally accessible \cite{SmerziRMP}. While in principle our approach generalizes to higher order non-linearities, we focus here on investigating the asymptoytic precision scaling these states can afford for frequency estimation with {\em quadratic} signal encoding, in the presence of  collective dephasing with arbitrary temporal correlations. In this way, both the important case of unentangled (CSS) initial states or the temporally uncorrelated Markovian regime may be recovered as limiting instances in our analysis.

After providing relevant background in Sec.\,\ref{sec:back}, our core results are presented in Sec.\,\ref{sec:quadratic}. By leveraging phase-space methods to describe the noisy dynamics, we provide analytical expressions for the relevant quantum Fisher information and explicitly determine the optimal measurement protocol that saturates the {\em quantum Cram\'{e}r-Rao bound} (QCRB) on precision. In the absence of temporal correlations, we find that OATS cannot outperform the linear $N^{-1}$ HL (or, equivalently, the generalized SQL for $k=2$). 
On the other hand, if dephasing is temporally correlated and a short-time ``Zeno regime'' is accessible \cite{Macieszczak}, our analysis shows that a scaling advantage over the linear HL may be achieved even for initial CSSs, of the form $N^{-5/4}$. Further to that, the use of squeezing, in combination with an interaction-based readout based on an ``echo protocol'' \cite{Monika,VladanSatin}, allows for the same optimal precision scaling of $N^{-3/2}$ that a maximally entangled generalized-GHZ state would attain, while avoiding the latter's fragility to preparation errors. In Sec.\,\ref{ratio}, we address the issue of noise-induced bias in local quadratic  estimation, and show how a simple {\em ratio estimator} which is asymptotically unbiased in the presence of dephasing noise may be constructed through the measurement of two distinct system observables. We prove that, at the cost of doubling the measurement resource overhead, this estimator still reaches the optimal precision mandated by the QCRB, as the standard estimator does.  We close in Sec.\,\ref{conclusion} with a summary of our main findings and a discussion of outstanding research questions. A number of appendices are included to both discuss possible physical implementations and to provide detailed derivations of all the results presented in the main text.

\section{Background}
\label{sec:back}

\subsection{Open-quantum system setting}

We consider $N\equiv 2 J$ qubit probes, whose Hamiltonian $H$ depends upon the target parameter $b$, in the presence of a collective (permutation-invariant) dephasing environment. While the case of a general nonclassical (e.g., bosonic\cite{FelixPRA,FUR,Bias}) environment is also of interest, in order to most clearly contrast the linear vs.\,non-linear noisy metrology setting, we focus here on the simpler case of a {\em classical} environment, which we model in terms of a (real) stochastic process $\xi(t)$. In units where $\hbar=1$, the noisy evolution is thus generated by
\begin{align}
H (t)= b \,J_z^k + J_z \, \xi(t)\, , \quad J_z\equiv \tfrac{1}{2} \sum_{\ell=1}^N \sigma_z^{(\ell)}, 
\label{Ham}
\end{align}
with $J_u, u\in \{u,y,z\}$ denoting total-spin angular-momentum operators, $\bf{\hat{z}}$ defining the quantization axis and $k \in \mathbb{N}$ the order of nonlinearity \footnote{It is immediate to see that noisy dynamics as in Eq.\,\eqref{Ham} violate the ``Hamiltonian-not-in-Kraus-span'' condition that is necessary for quantum error correction approaches to be viable, \cite{Layden2,Sisi}}. The well-established linear metrology scenario, whereby the probes are not allowed to interact following initialization, corresponds to $k=1$, whereas $k=2$ sets a quadratic coupling to the signal, which is the main focus of this paper. Although our theoretical analysis is implementation-independent, the above Hamiltonian can be mapped to different setups involving physical spins or atomic (pseudo-spin) degrees of freedom. In particular, $H(t)$ maps to a special limit of a BEC in the two-mode Josephson approximation (see Appendix \ref{app:BJJ}), which has been extensively studied as a setup for nonlinear sensing. In such a case, the frequency $b$ is an {\em internal} interaction-strength parameter related to a combination of atomic scattering lengths, whereas collective dephasing arises from both fluctuations in electromagnetic fields and atomic collisions, and is a leading source of decoherence. Connections to a platform capable of measuring an {\em external} field are less straightforward; in existing schemes \cite{Generating}, generation of the quadratic nonlinearity involves the measurement of an ancillary system, which results in a non-deterministic final state of the probe (see Appendix \ref{app:gen}). Higher-order nonlinearities, $k>2$ have also been studied \cite{Braunstein2008,DelCampo,Deng2021} but, because they entail $k$-body interactions, they remain challenging to implement.

Assuming that the stochastic process $\xi(t)$ in Eq.\,\eqref{Ham} is stationary, zero-mean, and Gaussian, the statistical properties of the noise are completely described by the two-time correlation function, $$C(t_2,t_1) \equiv \langle \xi(t_2)\xi(t_1) \rangle_{\xi} = \langle \xi(t)\xi(0) \rangle_{\xi} \equiv C(t),$$ where $\langle\bullet\rangle_{\xi}$ denotes an ensemble average over stochastic realizations and we have used time-translation-invariance to rewrite the correlator as a function only of the time lag, $t\equiv t_2-t_1 \geq 0$. Besides the full spatial correlations induced by the Hamiltonian in Eq.\,(\ref{Ham}), the resulting noise process then also exhibits in general non-trivial temporal correlations, which translates into a ``colored'' frequency spectrum,  
$$ S(\omega) \equiv \int_{- \infty}^{\infty} \!dt \,e^{- i \omega t} \langle \xi(t) \xi(0) \rangle_{\xi} =S(-\omega).$$ 
The case of Markovian, ``$\delta$-correlated'' noise, $C(t) \propto  \delta(t)$, is included as a limiting case.

Let $\{| J, m \rangle_{\bf{\hat{z}}} \}, \, -J \leq m \leq J$, denote the usual basis of {Dicke states}, namely, eigenstates of collective angular-momentum operators, with $J_z |J, m \rangle_{\bf{\hat{z}}} = m |J,m \rangle_{\bf{\hat{z}}}$, $J^2|J, m\rangle_{\bf{\hat{z}}} = \mbox{J(J+1)} |J,m \rangle_{\bf{\hat{z}}}$. The time-evolved noise-averaged density operator of a permutationally invariant initial state, $\rho(0)\equiv \rho_0$, may then be expanded in the Dicke basis, 
$$\bar{\rho}(t) \equiv \langle \rho(t) \rangle_{\xi} = \sum_{m,m'} \,_{\bf{\hat{z}}} \langle J, m |\bar{\rho}(t) |J, m'\rangle_{\bf{\hat{z}}}\; |J, m \rangle_{\bf{\hat{z}}}\, _{\bf{\hat{z}}} \langle J m'|, $$ 
where the time-dependent matrix elements may be expressed in the following form \cite{FUR}:
\begin{eqnarray}
&\, _{\bf{\hat{z}}}\langle J, m |\bar{\rho}(t) |J, m'\rangle_{\bf{\hat{z}}} = e^{i \varphi(t) }e^{- \gamma(t)}\, _{\bf{\hat{z}}}\langle J, m |\rho_0|J, m' \rangle_{\bf{\hat{z}}},\nonumber\\
&\:\varphi(t)= b t (m^2-m'^2), \quad \gamma(t)= \kappa(t) (m-m')^2.  \nonumber
\end{eqnarray}
Here, the phase factor $\varphi(t)$ gives the {\em quadratic signal encoding}, while the noise-induced decay factor $\gamma(t)$ captures the suppression of off-diagonal, coherence matrix elements. In turn, $\gamma(t)$ is a product of a state-dependent component and a {dynamic coefficient}, which is determined by an overlap integral of the noise frequency spectrum with the ``filter function'' describing, in this case, free evolution  \cite{FUR}:
\begin{eqnarray}
\kappa(t)= \frac{1}{32 \pi }\;  \!\int_{- \infty}^{\infty} \!\! d\omega\, \frac{\sin^2(\omega t/2)}{\omega^2} \,S(\omega).
\label{freqchi}
\end{eqnarray}
In the limit of temporally uncorrelated, Markovian noise is spectrum is flat, say, $S(\omega) \propto \gamma$, for some constant $\gamma >0$ with units of frequency, the above yields leading a linear temporal behavior, $\kappa(t) = \gamma \, t$. For temporally correlated noise instead, under the assumption that the spectrum decays to zero sufficiently rapidly above a high-frequency cutoff, say, $S(\omega) \approx 0$ for $|\omega| \gtrsim \omega_c$, Eq.\,\eqref{freqchi} implies a quadratic dependence upon time, say, $\kappa(t)$, $\kappa(t) \approx \kappa_0^2\, (\omega_c t)^2$, in the short-time regime $\omega_c t\ll1$, also referred to as the {\em Zeno limit} in the context of frequency estimation with non-Markovian environments \cite{Chin2012,Macieszczak,QuasiStatic}.

\subsection{Quantum estimation setting and noiseless precision bounds for quadratic metrology}

\subsubsection{Basics of parameter estimation}

Given the noisy Hamiltonian in Eq.\,(\ref{Ham}), and some specified {\em resource constraints}, our goal is to accurately estimate the frequency parameter $b$ as precisely as possible by implementing a suitable quantum metrology protocol. Here, we work under the assumption that both the number of sensors, $N$, and the {\em total} interrogation time, $T$, are fixed, with the number of experimental shots, $\nu \gg 1$, being finite but sufficiently large to ignore the effects of finite statistics. We further consider a \emph{local estimation} setting, whereby we aim to estimate a small deviation $\delta b \equiv b - b_0$ of the target parameter from a known value $b_0$, that we refer to as the \emph{operating point} and can be tuned to adjust the sensitivity of the measurement \cite{Degen}. In a regime where we have high prior knowledge, the deviation parameter $\delta b$ can be adjusted to be close to zero. A metrological protocol is then characterized by the initial state $\rho_0$, the encoding time $\tau \equiv T/\nu$ during which the signal imprints itself in the sensors' dynamics, and the positive operator-valued measurement (POVM) that is used for readout. Together, these features univocally determine the probability of measuring a string of outcomes $\vec{\mu}\equiv (\mu_1,\ldots,\mu_\nu)$, conditioned on the true value of the parameter being $b$, ${\mathbb P}(\vec{\mu}|b)$.

An estimate of the target parameter is inferred from the statistics of outcomes. An {\em estimator} $\hat{b}(\vec{\mu})$ is a function that associates each set of outcomes $\vec{\mu}$ with an estimate $\hat{b}$ of $b$. Being a function of random outcomes, the properties of $\hat{b}$ can be characterized in terms of its statistical moments. In particular, the mean and variance are, respectively, given by 
\begin{eqnarray*}
&& E[\hat{b}]
\equiv  \sum_{\vec{\mu}} {\mathbb P}(\vec{\mu}|b) \hat{b}(\vec{\mu}) = \langle \hat{b} (\vec{\mu}) \rangle_{\vec{\mu}}, \\
&&\Delta \hat{b}^2 \equiv \sum_{\vec{\mu}} {\mathbb P}(\vec{\mu}|b) \big( \hat{b}(\vec{\mu}) - \langle \hat{b}(\vec{\mu})_{\vec{\mu}} \big)^2 = \!
\langle \hat{b}(\vec{\mu})\rangle_{\vec{\mu}}^2  - \langle \hat{b}(\vec{\mu})\rangle^2_{\vec{\mu}} ,  
\end{eqnarray*}
where the expected value $\langle \bullet \rangle_{\vec{\mu}}$ is taken over {\em all} possible measurement outcomes and the pre-measurement state $\bar{\rho}(\tau)$. When the mean coincides with the true value, $E[\hat{b}] = b$, the estimator is said to be {\em unbiased}; otherwise, the difference $B[\hat{b}] \equiv E[\hat{b}]-b$ is the bias. A bias prevents us from obtaining an accurate estimate of the parameters true value, and for this reason it must be either eliminated or accounted for. The standard deviation, $\Delta \hat{b}\equiv (\Delta \hat{b})^{1/2}$, may be taken to quantify the precision of the estimate. 

The maximum amount of information that can be extracted from ${\mathbb P}(\vec{\mu}|b)$ is quantified in terms of the  {\em classical Fisher information} (FI), $F_{ \mathrm{cl}} [{\mathbb P}(\vec{\mu}|b)] \equiv \sum_{\vec{\mu}} (\mathbb P(\vec{\mu}|b))^{-1} ( \partial_b \mathbb P(\vec{\mu}|b))^2 $. Maximizing the FI over all possible POVM measurements that quantum mechanics allows leads to the {\em quantum FI} (QFI), $F_{ \mathrm{Q}} [\bar{\rho}(\tau) ]$. Provided that the estimator is unbiased, an ultimate lower bound to the achievable precision is set by the {\em quantum Cram\'{e}r-Rao bound} (QCRB) \cite{Holevo,Helstrom}, 
\begin{equation}
\Delta \hat{b}^2(\tau) \geq \Delta \hat{b}^2_{\mathrm{QCR}}(\tau) =( {\nu F_{ \mathrm{Q}} [\bar{\rho}(\tau)] })^{-1}. 
\label{qcr}
\end{equation} 
In the common scenario where detection is implemented as a projective readout of an observable $\mathcal{O}$, an estimator $\hat{b}$ may be constructed from the sample mean, $\langle \hat{\mathcal{O}}(\tau)\rangle = \nu^{-1} \sum_{i=1}^{\nu} \mu_i$, by inverting the functional dependence of the expectation value upon the target parameter. If $\langle \mathcal{O}(\tau) \rangle \equiv f(b;\tau)$, we may assume $\langle \mathcal{O}(\tau) \rangle \approx \langle \hat{\mathcal{O}}(\tau)\rangle \equiv \hat{f}(b;\tau)$ in the asymptotic regime where $\nu\gg1$; thus, $b= f^{-1}(\hat{f}(b;\tau))$, provided that the relationship is {\em one-to-one} throughout a neighborhood of the operating point. The above procedure constitutes the {\em method of moments} \cite{SmerziRMP}. Explicitly, an expression for the uncertainty is obtained by observing that, for $\nu \gg1$, we may equate 
$$ \hat{b}-b_0 = [\partial_b f(\hat{b}){\vert_{b_0}}]^{-1} 
\big[ \langle \hat{\mathcal{O}}(\tau)\rangle - \langle \mathcal{O} (\tau) \rangle_{b_0}\big], $$
and by using error propagation to arrive at \cite{FUR,Bias}
\begin{eqnarray}
\Delta \hat{b}^2(\tau)  = \frac{\Delta \mathcal{\hat{O}}^2(\tau) }{
\big[\partial_b \langle \mathcal{O}(\tau) \rangle {\vert_{b_0}} \big]^{2} }
 =  \frac{\Delta \mathcal{O}^2(\tau) }{
 \nu \big[\partial_b \langle \mathcal{O}(\tau) \rangle {\vert_{b_0}} \big]^{2}} ,
\label{mom} 
\end{eqnarray}
with $\Delta \mathcal{\hat{O}}^2(\tau)\equiv [\langle \mathcal{O}^2(\tau)\rangle - \langle \mathcal{O}(\tau)\rangle^2 ]\nu^{-1} $ being the sample-mean variance \cite{erratum}.
On the other hand, the QCRB in Eq.\,\eqref{qcr} may be saturated by a POVM that projects onto the basis that diagonalizes the so-called {\em symmetric logarithmic derivative} (SLD) operator \cite{Pezze2014}, $\hat{L}$, which satisfies $F_Q[\bar{\rho}(\tau)]= \text{Tr}[\bar{\rho}(\tau) \hat{L}^2 ]$. In a noiseless setting, the QFI of a {\em pure} state $\rho(\tau) \equiv |\psi (\tau)\rangle\langle \psi(\tau)|$ encoded in the evolution of $|\psi_0\rangle$ under the Hamiltonian $H_b \equiv  b H = b J_z^k$ takes the simple, $b$-independent form $F_{ \mathrm{Q}} [\rho(\tau) ] = 4 \nu \,\Delta H(\tau)^2$. While this parameter independence is preserved so long as the signal remains encoded unitarily, the presence of noise usually hinders an exact computation of the QFI. Notably, in the next section we will show how the QFI can be analytically computed in the asymptotic regime $N \gg 1$ for a class of metrologically relevant states that lead to a gain in precision scaling with respect to the linear HL.

In general, the presence of noise as in Hamiltonian Eq.\,\eqref{Ham} does not only degrade the precision, but also introduce noise-dependent \emph{bias} $B_\xi [\hat{b}] $ in the estimator $\hat{b}$ we discussed in the context of moment-based local estimation. This bias is present even in the limit of infinite measurements, forcing us to reassess the procedure used to evaluate $\hat{b}$ for a noiseless system. In prior work \cite{FUR}, we have addressed this problem for the standard setting of linear frequency estimation, $k=1$. In such a case, it was shown that an asymptotically unbiased estimator reaching the {\em same} precision as the QCRB can be constructed by measuring \emph{two} distinct system observables. In Sec.\,\ref{ratio} we extend these results to quadratic metrology. 

\smallskip

\subsubsection{Noiseless precision bounds}
\label{nsless}

The optimal performance bounds for metrology with quadratic encoding in the absence of noise have been rigorously derived in the asymptotic regime of large $N$ for both classical and (maximally) entangled input states\cite{DattaPRA} (see also Appendix \ref{app:noiseless} for an explicit derivation). By letting $k=2$ and $\xi(t)\equiv 0$ in Eq.\,(\ref{Ham}), the best classical preparation is found within the set $\mathcal{C}_{\bf{\hat{n}}}(\theta, \phi)=\{ \rho_{\bf{\hat{n}}}\}$ of CSS. Here, $\rho_{\bf{\hat{n}}}$ is the density operator corresponding to $|{\rm CSS}\rangle_{\bf{\hat{n}}}$, which obeys $J_{\bf{\hat{n}}} |{\rm CSS}\rangle_{\bf{\hat{n}}}= J|{\rm CSS}\rangle_{\bf{\hat{n}}}$, with $J_{\bf{\hat{n}}} \equiv \cos{\phi}\,\sin(\theta)\, J_x + \sin(\phi)\sin(\theta)\, J_y + \cos(\theta)\, J_z$ being the spin component along direction $\bf{\hat{n}}$ in the Bloch sphere, and $0 \leq \theta \leq \pi $, $ 0 \leq \phi \leq 2\pi$. Importantly, since the Hamiltonian in Eq.\,(\ref{Ham}) has a manifest symmetry along the $\bf{\hat{z}}$ axis, we may fix $\phi=0$ and restrict the initial set of states to the subset $\mathcal{C}_{\bf{\hat{q}}}(\theta)$, with $\bf{\hat{q}}\equiv \cos(\theta) \bf{\hat{x}}+ \sin(\theta) \bf{\hat{z}} $ being the $\theta$-dependent direction. Alternatively, note that we may characterize $\mathcal{C}_{\bf{\hat{q}}}(\theta)$ as the set of states we obtain by rotating  $\rho_{\bf{\hat{z}}}$, 
$$\mathcal{C}_{\bf{\hat{q}}}(\theta)= \{ \rho_{\bf{\hat{q}}} (\theta) \,:\,\rho_{\bf{\hat{q}}}(\theta) = \mathcal{R}(\theta)\rho_{\bf{\hat{z}}}\, \mathcal{R}^{\dagger}(\theta) \},$$   
with $\mathcal{R}(\theta)\equiv e^{i (\theta-\pi/2) J_y}.$ For the optimal choice of  $\theta=\pi/4$, applying Eq.\,(\ref{mom}), a measurement of $J_y$ yields uncertainty 
\begin{equation}
\Delta \hat{b}_{\text{opt}}^{\text{CSS,0}}(\tau) = 2\,(T\tau)^{-1/2} N^{-3/2}, \quad N\gg1. 
\label{quadrsql}
\end{equation}
It is interesting to note that, unlike for linear metrology where the ideal CSS is orthogonal to the signal direction ($\theta= 0$), here it lies halfway through the $\bf{\hat{x}}$-$\bf{\hat{z}}$ plane. The same precision may be reached by any CSS with $\phi \neq 0$ by rotating the measurement basis, $J_y \mapsto J_y \cos(\phi) + J_x \sin(\phi)$.

Among nonclassical states, an $N$-partite entangled, generalized GHZ state is known to saturate the QCRB. That is, two states corresponding to the maximum and minimum eigenvalue of $J_z^2$ are superposed, yielding a ``$\Phi$ state'' of the form $|{\rm \Phi}\rangle\equiv (|J,J \rangle_{\bf{\hat{z}}}\!+|J,0\rangle_{\bf{\hat{z}}})/\sqrt2$. Computation of the noiseless QFI then leads, through Eq.\,(\ref{qcr}), to the bound on precision, 
\begin{equation}
\Delta \hat{b}_{\text{opt}}^{\Phi,\text{0}} (\tau) =4\, (T \tau)^{-1/2} N^{-2}, \quad N\gg1.
\label{quadrhl}
\end{equation}
This can be saturated by implementing a measurement of the survival probability, $\mathcal{O}_{|\Phi\rangle} \equiv |\Phi\rangle \langle \Phi| -(I-|\Phi\rangle \langle \Phi|)$. For both kinds of states, note the absence of an optimal encoding time as $\Delta \hat{b}(\tau)$ is monotonically decreasing in $\tau$. Eqs.\,\eqref{quadrsql} and \eqref{quadrhl} may be taken to define a ``generalized SQL'' and a ``generalized HL'', respectively, as appropriate for quadratic metrology.

\section{Quadratic signal encoding under collective dephasing}
\label{sec:quadratic}

\subsection{Noisy precision bounds for product and maximally entangled initial states}
\label{noisyb}

Having established the strategies leading to optimal precision scaling using classical and (maximally) entangled input states in the absence of noise, a natural question is how their performance is degraded in the presence of collective dephasing. With that in mind, let us again consider an optimal noiseless CSS input state, with $\theta=\pi/4, \phi=0$, now evolving under Hamiltonian (\ref{Ham}) with $\xi(t) \neq 0$, after which a measurement of $J_y$ is implemented. In this case, the presence of dephasing makes long encoding times disadvantageous. Instead, a measurement time $\tau_{\text{opt}}^{\text{CSS}}$ optimizing precision emerges from the interplay between the decoherence buildup and the signal accumulation. Despite the noisy dynamics, exact formulas for the relevant mean values $\langle J_y(t)^n\rangle$, $n \in \{1,2\}$, entering Eq.\,(\ref{mom}) can still be derived. A short-time analytic expansion of the ensuing expression for $\Delta \hat{b}(t)$, complemented with numerical optimization over the encoding period leads to the following bounds (see Appendix \ref{app:noisyphi} for detail):
\begin{eqnarray*}
\Delta \hat{b}^{\text{CSS}}_{\text{opt}} =2\, (\gamma/T)^{1/2} N^{-1}, \quad \text{Markovian noise},\;\; \\
\;\;\Delta \hat{b}^{\text{CSS}}_{\text{opt}} = 2^{3/2}(\kappa_0 \omega_c/T)^{1/2}\, N^{-5/4},  \quad \text{Zeno regime}.
\end{eqnarray*}

When preparing the system in the entangled state $|{\rm \Phi}\rangle$, the time-evolved density operator remains supported on a two-dimensional subspace in the presence of collective dephasing. This still allows, exceptionally, for an exact evaluation of the QFI, as detailed again in Appendix \ref{app:noisyphi}. Optimizing with respect to time leads now, via the QCRB, to the bounds 
\begin{eqnarray*}
\quad\Delta \hat{b}_{\text{opt}}^{\Phi} = 2 (2e)^{1/2}\,(\gamma/T)^{1/2}\, N^{-1}, \quad \text{Markovian noise},\\
\Delta \hat{b}^{\Phi}_{\text{opt}} = 4 e^{1/4} (\kappa_0 \omega_c/T)^{1/2}\,N^{-3/2}, \quad \text{Zeno regime},\;\,
\end{eqnarray*}
which, as in the noiseless limit, can be achieved by implementing a measurement of the survival probability. 

The above bounds indicate that, similar to linear metrology, collective dephasing is far more harmful than spatially uncorrelated one. In the Markovian limit, the precision is degraded below the generalized SQL of $N^{-3/2}$ that is known to hold for uncorrelated dephasing \cite{DelCampo}. Notably, however, a scaling at the (linear) HL is maintained, with a classical CSS outperforming the entangled $\Phi$ state by a constant factor. Also similar to linear metrology, temporally correlated noise is less detrimental than its Markovian counterpart. By comparing the two bounds above in the Zeno regime, entangled states are seen to offer a metrological scaling advantage over product states, and super-Heisenberg precision is now possible, with respect to the linear $1/N$.  The optimal scaling, nevertheless, is still bound to the quadratic SQL, suggesting that a {\em generalized no-go} for sub-SQL precision may still apply under collective dephasing noise, as in the linear setting \cite{third}.

\subsection{Noisy precision bounds for one-axis-twisted states}

\subsubsection{Holstein-Primakoff description of the dynamics}
\label{HPpres}

While important on theoretical grounds, computing the asymptotic performance of the $\Phi$ state is of little practical value, as the generation of $N$-partite entangled GHZ states for large $N$ remains very challenging. However, the nonclassical OATSs, which include CSS as a particular limit, are accessible in current atomic interferometers \cite{SmerziRMP} and capable of breaking the SQL in the linear setting \cite{Monika,VladanSatin}, while also offering more robustness against preparation errors \cite{SchulteEchoes}. This motivates us to explore their potential metrological relevance in the noisy quadratic encoding setting as well. However, multiparticle entanglement makes the evolution of an initial OATS highly non-trivial \cite{FelixPRA}, and the mean values that are needed to obtain the uncertainty via the method of moments [Eq.\,(\ref{mom})] are no longer exactly computable even in the simplest case where $\mathcal{O}$ is a collective spin observable. For this reason, a more sophisticated approach for computing the asymptotic precision scaling is needed. 

The line of attack we follow here builds on an approach introduced by Knysh {\em et al.}\cite{Durkin} to derive precision bounds in a linear phase-estimation context. It relies on the well-known Holstein-Primakoff (HP) transformation \cite{HP}, which maps spin observables into creation and annihilation operators $a,a^{\dagger}$ obeying bosonic commutation rules, $[a^{\dagger},a]= I$, specifically:
\begin{eqnarray*}
 &&J^+ \equiv \sqrt{2J} \sqrt{1- \tfrac{a^{\dagger}a}{2J} }\, a \approx \sqrt{2J} \,a, \\
 && J^- \equiv \sqrt{2J} a^\dag \sqrt{1- \tfrac{a^{\dagger}a}{2J} }  \approx \sqrt{2J} \,a^{\dagger}, \quad J_z \equiv J-  a^{\dagger} a, 
\end{eqnarray*}  
where the approximate equalities hold in a \textit{small excitation regime}, such that ${a^{\dagger}a}/(2J) \ll 1$. Inverting the above relationship makes it then possible to approximately express the angular momentum components in terms of position and momentum quadratures, namely, $ J_x \approx \hat{x}, J_y \approx J \hat{p}, J_z \approx J$.

We restrict the initial states of interest to the set $ \mathcal{G}_{\bf{\hat{z}}} \equiv \{\rho_{\bf{\hat{z}}}(\mu, \beta)\}$, with $\rho_{\bf{\hat{z}}}(\mu, \beta)$ being the density operator of a pure OATS generated by first squeezing a $|{\rm CSS}\rangle_{\bf{\hat{z}}}$ with a fixed direction ($\theta=0$), and then applying a rotation that {effectively rearranges the $\hat{x}$ and $\hat{p}$ quadratures}; that is, 
\begin{equation}
|\text{OATS}\rangle_{\bf{\hat{z}}} \equiv e^{-i \beta J_z} e^{-i \mu J_x^2}  |\text{CSS}\rangle_{\bf{\hat{z}}}.
\label{oatz}
\end{equation}
Note that, in the above HP regime, $|\text{CSS}\rangle_{\bf{\hat{z}}}$ maps to the ground state $|0\rangle$ of the harmonic-oscillator Hamiltonian $a^{\dagger}a$, which has precisely zero excitations. To reintroduce a dependence upon the direction $\theta$ of the initial CSS, we allow for the direction of the operator that encodes the signal to vary, $J_{\bf{\hat{z}}} \mapsto J_{\bf{\hat{m}}}$, with {$\bf{\hat{m}}\equiv \sin(\theta) \bf{\hat{x}}+ \cos(\theta) \bf{\hat{z}}$.} 
In this way, by observing that $\mathcal{O}(\hat{x} ,\theta)= \sin(\theta) \hat{x}  + J \cos(\theta) $ is the HP limit of $J_{\bf{\hat{m}}}$ for $J \gg 1$, the noisy dynamics may be written as
\begin{eqnarray}
H(\theta, t) \approx\, b\,\mathcal{O}(\hat{x},\theta)^2  + \mathcal{O}(\hat{x},\theta) \xi(t).
\label{HHP}
\end{eqnarray}
This is equivalent to a setting where the quantization axis remains fixed, as assumed in Eq.\,(\ref{Ham}), but the direction of the initial OATS depends on $\theta$, so that the initial states of interest belong to 
$${\mathcal{G}_{\bf{\hat{q}}}(\theta) \equiv \big\{ \rho_{\bf{\hat{q}}}(\mu, \beta, \theta): \rho_{\bf{\hat{q}}}(\mu, \beta,\theta) =\!\mathcal{R}(\theta)\rho_{\bf{\hat{z}}}(\mu, \beta) \mathcal{R}^{\dagger}(\theta)\big \}},$$
where, as in Sec.\,\ref{nsless}, $\mathcal{R}(\theta) = e^{i (\theta- \pi/2) J_y}$ and $\bf{\hat{q}}\equiv \cos(\theta) \bf{\hat{x}}+  \sin(\theta) \bf{\hat{z}}$ (see Appendix \ref{equivalence} for further details).

\begin{figure*}[t!]
\centering
\includegraphics[width=18cm]{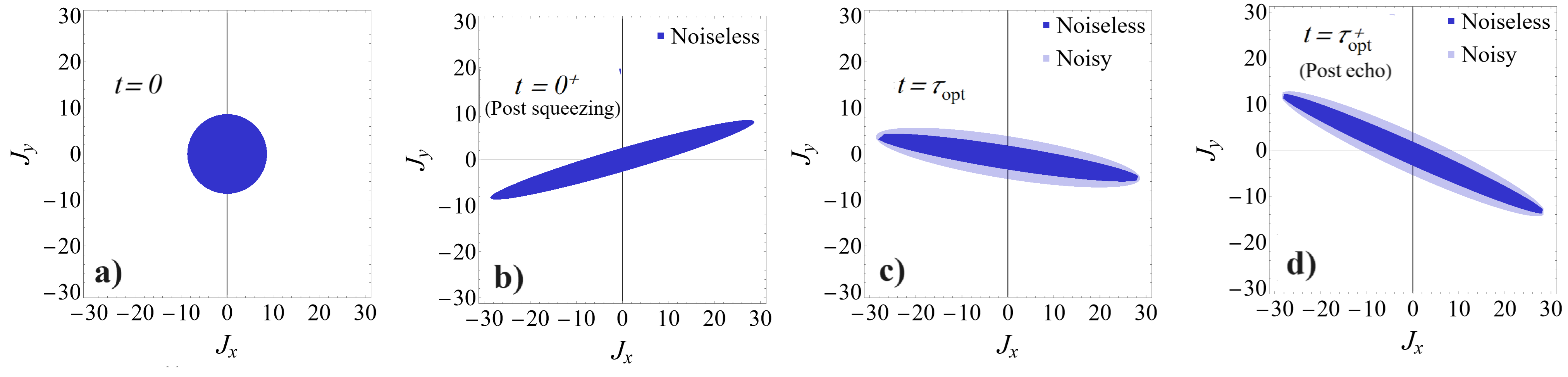}
 \vspace*{-5mm}
\caption{{\bf Time evolution under collective dephasing in phase space.} From left to right, snapshots are shown of the Wigner function in the HP representation, in the small excitation regime, with darker vs.\,lighter (blue) color corresponding to noiseless vs.\,noisy dynamics.  
Panel a): Initial CSS state, $|\text{CSS}\rangle_{\bf{\hat{z}}} \simeq|0\rangle$.
Panel b): |OATS$\rangle_{\bf{\hat{z}}} \simeq e^{-i \beta (J-a^{\dagger}a)} e^{-i \mu \hat{x}^2}|0\rangle$, right after a rotation and a squeezing operation are implemented, Eq.\,\eqref{oatz}.
In the noisy case, diffusion along the $J_y$-direction increases uncertainty. 
Panel c): Time-evolved state at time $\tau_{\text{opt}}$ [Eq.\,\eqref{optpar}], at which a phase $\varphi_\text{opt}=b \tau_{\text{opt}}$ is imprinted.
Panel d): State right after an anti-squeezing operation $e^{i \eta \hat{x}^2}$ is applied, where the strength $\eta$ depends on the initial state parameters $\{\mu, \beta\}$ as in Eq.\,(\ref{eta}). This gives the pre-measurement state for readout, see also Eq.\,\eqref{readout}.  Note the vertical displacement of $W_{\rm OATS}(x,p,t)$, which allows for signal detection by measuring $J_y$. Here, $N=10$, $\mu=1/\sqrt{2J}$, and $\beta=-\pi/2 $, which lead to optimal precision scaling within the members of $\mathcal{K}_{\bf{\hat{z}}}$, as shown in Table \ref{table:1}. We consider temporally correlated dephasing in the Zeno regime, $\kappa(t) \approx \kappa_0^2 (\omega_c t)^2$, with $\kappa_0^2= 6$, $\omega_c=1$, and $\tau_{\text{opt}} \approx 0.19$.
}
\label{Wigners} 
\end{figure*}

The key observation to proceed is to note that, for every fixed noise realization, the evolution due to the Hamiltonian in Eq.\,(\ref{HHP}) can be decomposed into the product of a signal-encoding unitary, $U_{\rm S}(\theta,t)\equiv e^{-i b t\,\mathcal{O}(\hat{x},\theta)^2 }$, and a random but still unitary phase evolution, $U_{\xi}(\theta, t)= e^{-i \mathcal{O}(\hat{x}, \theta) \int_0^t\!ds\, \xi(s)}$. Let $\rho_0 =|\psi\rangle \langle \psi|$, with  $|\psi\rangle \simeq  
e^{-i \beta (J-a^{\dagger} a)} e^{-i \mu \hat{x}^2} |0\rangle$ the HP limit of $|\text{OATS}\rangle_{\bf{\hat{z}}}$, be the initial state. Then, the ensemble-averaged effect of the noise $\xi(t)$ on the time-evolved state may be obtained by performing a cumulant expansion which, thanks to the assumed Gaussian statistics, truncates {\em exactly} to the second order. We may thus write
\begin{eqnarray}
     \left \langle \langle x | U_{\xi}(t)\rho_0 U^{\dagger}_{\xi}(t) |y \rangle \right \rangle_{\xi} & = & \langle x |
\rho_0  |y \rangle  e^{- \kappa(t) \sin(\theta)^2 (x-y)^2} \notag \\
& \equiv & \langle x | \rho_0  |y \rangle  e^{-\mathcal{T}_{xy}(\theta, t) }  ,
\label{Gauss} 
\end{eqnarray}
where, in the second line, we have introduced a position-dependent dephasing factor, $\mathcal{T}_{xy}(\theta, t)$. It follows that the density operator may be represented in the position eigenbasis as 
\begin{equation}
\bar{\rho}(t) =\iint_{-\infty}^{\infty}\! dx\, dy\, \langle  x |\rho_0| y \rangle  \mathcal{T}_{xy}(\theta, t) \,
U_{\rm S}(\theta, t) |x\rangle \langle y| U_{\rm S}^{\dagger}(\theta, t) .
\label{eqev0}
\end{equation}
Taking the time derivative immediately leads to the following time-local master equation for the density operator:
\begin{equation}
\dot{\bar{\rho}}(t)= - i b\, [ \mathcal{O}(\hat{x},\theta)^2,\bar{\rho}(t)] {- \frac{1}{2} \dot{\kappa}(t) \sin(\theta)^2 \Big [ x, [x, \bar{\rho}(t) ]\Big ] .}
\label{ME}
\end{equation}
Formally, a solution for the noise-averaged density operator may be obtained as the product of position and momentum operators. The key idea is that $\mathcal{T}_{xy}$ can be regarded as a thermal density matrix element of a freely evolving fictitious particle of mass $2 \kappa(t) \sin(\theta)$  written in the position eigenbasis:
\begin{eqnarray*}
\mathcal{T}_{xy} (\theta, t) = \frac{\langle x| e^{- \hat{p}^2/(4 \kappa(t) \sin(\theta))} |y\rangle}{\sqrt{ 4 \pi \kappa(t) \sin(\theta)}}  
\equiv  \langle x| \mathcal{T}(\hat{p},\theta, t)| y \rangle.
\end{eqnarray*} 
Upon defining the input-state-dependent operator
\begin{eqnarray*}
\mathcal{V}_{|\psi\rangle } (\hat{x},\theta, t)\equiv U_{\rm S}(\theta, t) \int_{-\infty}^{\infty}\! dx\, \langle x| \psi \rangle\, |x\rangle \langle x| , 
\end{eqnarray*}
which is diagonal in the position eigenbasis, we can then check by direct comparison with Eq.\,(\ref{eqev0}) that the solution $\bar{\rho}(t)$ may be compactly expressed as 
$$\bar{\rho}(t)=\mathcal{V}_{|\psi\rangle }(\hat{x},\theta, t) \mathcal{T}(\hat{p},\theta, t) \mathcal{V}_{|\psi\rangle }^{\dagger}(\hat{x},\theta, t).$$ This generalizes the construction of  Ref.\,\onlinecite{Durkin} to cases where the signal and OATS direction, $\bf{\hat{m}}$ and $\bf{\hat{z}}$, are not perpendicular.

Moving forward, it is convenient to tackle the system dynamics in the phase-space representation (see also Appendix \ref{app:phasespace}). Its cornerstone, the {\em Weyl transform}, assigns a function $f_{\mathcal{A}}(x,p)$, depending on classical position and momentum coordinates $x$ and $p$, to every operator $\mathcal{A}$, via the mapping  
$$f_{\mathcal{A}}(x,p)  \equiv \frac{1}{\pi} \int_{-\infty}^{\infty}\, dy \, \langle x + y | \,\mathcal{A}\, | x-y \rangle\, e^{-2 i p y}.$$ 
In particular, the Weyl transform of a density operator $\rho(t)$ yields the Wigner function $W_{\rho}(x,p,t)$. Thus, 
the Wigner function of any state evolving from an initial condition in $\mathcal{K}_{\bf{\hat{z}}}$ can be computed by first expressing $W_{\rm OATS}(x,p,0)$ with the help of a Mehler's kernel, then using a Green's function approach to propagate forward in time. While we give full detail in Appendix \ref{app:compWig}, the resulting structure is remarkably simple, and corresponds to a displaced harmonic oscillator: 
\begin{equation}
W_{\rm OATS}(x,p,t) = \frac{1}{\sqrt{\pi^2 Q(\theta,t)}}\, e^{  - \frac{1}{J \delta} x^2 - \frac{J \delta}{Q(\theta,t)} (p+ 2 \eta x + \varphi\,u)^2},
\label{wig} 
\end{equation}
where $\varphi \equiv b\,t$ is the phase, $u \equiv 2\sin(\theta) (J \cos(\theta)+ x \sin(\theta))$, and the other relevant parameters are given by
\begin{eqnarray}
&&\eta \equiv 2 \mu  [\cos(2 \beta) + J \mu \sin(2 \beta)] (2  \delta)^{-1}, \label{eta} \\
&&\delta  \equiv \cos(\beta)^2 +(1+4 J^2 \mu^2) \sin(\beta)^2 + 2 J \mu \sin(2\beta),\qquad  \label{delta}\\
&&Q (\theta, t) \equiv  4  J \delta\, \kappa(t) \sin(\theta)^2+1 .
\label{Qt}
\end{eqnarray}
That is, the dynamics results in $J$-dependent effective displacement $\eta$ and quadrature squeezing $\delta$, with the influence of decoherence being captured by $Q(\theta, t)$. Eq.\,(\ref{wig}) allows us to visualize the evolution in phase-space that the Hamiltonian in  Eq.\,(\ref{HHP}) generates in this limit, the effect of collective dephasing is simply to induce {\em diffusion} along the $J_y$ axis, see also Fig.\,\ref{Wigners}. Importantly, however, restrictions must be imposed to ensure that the low-excitation condition is obeyed throughout the time regime of interest. As we explicitly show by expressing the mean excitation number $\langle a^\dag a\rangle$ in phase-space representation [see in particular Eq.\,\eqref{exc}], the subset of states $\mathcal{K}_{\bf{\hat{z}}}\subset \mathcal{G}_{\bf{\hat{z}}}$ that are adequately described by Eq.\,(\ref{HHP}) must have $0 \leq \mu \lesssim (2J)^{-1/2}$: in other words, they must be \emph{properly squeezed} \cite{SmerziRMP}, or Gaussian. Even so, as time evolution increases the value of $\langle a^{\dagger} a \rangle$, the validity of our description is restricted to sensing {\em small} phase fluctuations over the encoding period, $\varphi = b \tau \approx 0$, in the short-time regime of metrological interest -- that is, $\gamma \tau \ll 1$ or $\omega_c \tau \ll 1$, for Markovian or non-Markovian noise, respectively.

\begin{table*}[t]
 \centering
\begin{tabular}{||c| c c c c c ||} \hline
Initial & Squeezing &  Rotation  &  QCRB $[\left( \kappa_0 \omega_c/T\right)^{1/2}]$  & QCRB[$(\gamma/T)^{1/2}$] & \; QCRB $[(1/ \tau T)^{1/2}]$ 
 \\ [0.5ex] 
state & angle, $\mu$  & angle, $\beta$  & Non-Markovian Zeno limit  & Markovian noise & Noiseless limit \\ [0.5ex] 
 \hline\hline
$\rm CSS$ & 0 & $\pi/2$  &  $\left(6 \,\sqrt{3}\, \right)^{1/2}  N^{-5/4} $ & $2 \,N^{-1}$ & $2\,N^{-3/2}$\\
$\;\;\rm  KU\, OATS\;\;$ & $\;\;2\,3^{1/6}  N^{-2/3}$ & $\;\;\pi/2- 3^{-1/6}\,N^{-1/3}$ & $2^{-1/4} \,{3^{2/3}}  \, N^{-17/12}  $  & $2\,N^{-1}$  & $-$\\
 PE\,OATS & $N^{-1/2}$ & $-\pi/2$  & $(3^{5/4}/2^{3/4})\,N^{-3/2}$ & $2\,N^{-1}$  & $4\,N^{-2}$\\ 
 $|\Phi \rangle$ state &  $-$ & $-$ &  $4 \,e^{1/4}\,N^{-3/2}$ & $2\,\sqrt{2 \, e}\; N^{-1}$ & $4\,N^{-2}$ \\ [1ex] 
 \hline
\end{tabular}
\caption{
{\bf Asymptotic performance of generalized Ramsey protocols for quadratic signal encoding  under collective dephasing.} In the first three rows, the initial state is given by $|\text{OATS}\rangle= e^{-i \beta J_z} e^{-i \mu J_x^2}| \text{CSS}\rangle_{\bf{\hat{z}}}$, with squeezing and rotation angles $\mu$,  $\beta$ specified in the second and third columns. KU denotes the OATS with initial minimal $y$-dispersion considered by Kitagawa and Ueda \cite{Kita1993}, whereas PE denotes OATS used in echo-based protocols \cite{Monika}. The signal is encoded quadratically along an optimal direction ${\bf{\hat{m}}}_{\text{opt}}= \sqrt{1/3} \,{\bf{\hat{x}}} + \sqrt{2/3} \,{\bf{\hat{z}}} $, corresponding to $\theta_{\text{opt}}$ in Eq.\,\eqref{optpar}. The optimal POVM is realized by implementing a nonlinear readout as in Eq.\,\eqref{readout}, and saturates the time-optimized QCRB, for dephasing in the Zeno or Markovian regimes (fourth and fifth columns) and in the absence of noise (last column). Precision bounds for the $N$- entangled, non-Gaussian $\Phi$ state (fourth row) are also included; in this case, they may be reached by a measurement of the survival probability.}
\label{table:1}
\end{table*}

\subsubsection{Quantum Fisher information and optimal precision scaling}

Crucially, in addition to providing a useful visualization, the phase-space representation enables for an exact solution of the SLD operator that determines the QFI. While again we present the full details of the calculation elsewhere (Appendix \ref{app:ultimate}), the salient aspects may be summarized as follows. Consider the operator equation that implicitly defines the SLD, 
\begin{align}
   \frac{\partial}{\partial b}\bar{\rho}(t)=  - i\, t \big[\bar{\rho}(t), H (\theta, t)\big] \equiv \frac{1}{2} \big\{ \hat{L}(\theta,t), \bar{\rho}(t) \big\}, 
     \label{equality}
\end{align}
with $\bar{\rho}(t)$ being the time-evolved density operator of a member of $\mathcal{K}_{\bf{\hat{z}}}$ under the Hamiltonian in Eq.\,\eqref{HHP}, 
and the curly brakets denoting the anti-commutator. The main point is that, by rewriting Eq.\,(\ref{equality}) in the phase-space language, it becomes an {\em algebraic} equation which is easy to solve. To see this, we note that the Weyl transform of the left hand-side of Eq.\,(\ref{equality}) reduces to $W_{\bar{\rho}}(x,p,t)$ multiplied by a quadratic function of the position and momentum coordinates, thanks to the Gaussianity of the Wigner function in Eq.\,(\ref{wig}). Working at the operating point $b_0=0$, and for $f_{\mathcal{O}}(x,\theta) \equiv \sin(\theta)\, x\,+ J \cos(\theta)$ denoting the Weyl transform of $\mathcal{O}(\hat{x}, \theta)$, we have 
\begin{eqnarray}
  & - i\, t\, f_{ [\bar{\rho}(t), H ]}(x,p,\theta)= \big \{ 4 J \,\delta \,t \,Q(\theta, t)^{-1} \sin(\theta) 
  \label{LHS} \\
  & \hspace{12mm}\times \big[(p+ 2 \eta x) \, f_{\mathcal{O}}(\theta, x) \big] \,W_{\text{OATS}}(x,p,t) \big\}, \notag
\end{eqnarray}
where the noise enters as a multiplicative factor, independent of $x,p$, and $Q(\theta,t)$ is given in Eq.\,\eqref{Qt}. Upon proposing the Ansatz $\hat{L}\equiv  \sum_{q,j} \alpha^{(j)}_{q}\, q^{j}+ \beta \,\hat{x} \hat{p}$, with $q \in \{\hat{x},\hat{p}\}$, $j \in \{1,2\}$, and $\alpha^{(j)}_q$ and $\beta$ unknown, and substituting in the Weyl transform on the right hand-side of Eq.\,(\ref{equality}), we can solve for the undetermined coefficients by comparing with Eq.\,(\ref{LHS}) term-by-term. Transforming back to Hilbert space, the SLD is then found to be the sum of two distinct contributions, say, $\hat{L} \equiv -4 t \sin(\theta)(\hat{L}_A + \hat{L}_B)$, with
\begin{eqnarray}
\hat{L}_A(\theta, t)& \equiv  &  {\frac{\delta  J^2 \cos(\theta)}{Q(\theta, t)}}
\, \big(\hat{p}+ 2 \eta \hat{x}\big) ,
\label{LA} \\
\hat{L}_B (\theta, t) &\equiv & {\frac{\delta  J \sin(\theta) }{Q(\theta, t)+1}}
\, \Big(\frac{1}{2} \{\hat{x}, \hat{p}\} + 2 \eta \hat{x}^2 )\Big) .
\label{LB}
\end{eqnarray}
Interestingly, as $J \eta \ll 1$ for states in $\mathcal{K}_{\bf{\hat{z}}}$, {and recalling that $J_y \approx Jp$}, Eq.\,(\ref{LA}) may be interpreted as the HP limit of a \emph{nonlinear readout}, $\hat{L}_{A} \propto e^{-i \eta J_x^2} J_y e^{-i \eta J_x^2} $, which has been used to avoid single particle detection {in twisting-echo and time-reversal-based metrological protocols} \cite{Monika,VladanSatin}. Eq.\,(\ref{LB}), on the other hand, involves measurement of \emph{quadratic} observables $J_x^2$ and $\{J_x,J_y \}= (J_x+J_y)^2 - J_x^2 - J_y^2$, 
which may prove challenging to implement.

Computing the QFI as the SLD dispersion, $F_Q[\bar{\rho}(t)]= {\rm Tr}[ \bar{\rho}(t) \hat{L}^2]$, we find that the cross term $\text{Tr}[\{ \hat{L}_{A}, \hat{L}_B \} \bar{\rho}(t)]$ vanishes exactly for $\rho_0 \in \mathcal{K}_{\rm{\hat{z}}}$. It follows that the end result, $F_Q[\bar{\rho}(t)]\equiv 2 \delta  J^2 t T \sin(\theta)^2 \tilde{F}_Q[\bar{\rho}(t)]$, with
\begin{eqnarray}
\tilde{F}_{Q}[\bar{\rho}(t)] &=&  \frac{4 J \cos(\theta )^2}{
Q(\theta,t)}  + \frac{2 \delta  \sin ^2(\theta )}{Q(\theta, t)+1}  \\
&\equiv& \tilde{F}_{Q}^{(A)} [\bar{\rho}(t)]+ \tilde{F}_{Q}^{(B)} [\bar{\rho}(t)],
\label{FQB}
\end{eqnarray}
can be neatly separated into the sum of two terms, $\text{Tr}[ \hat{L}_{A}^2 \bar{\rho}(t)]$ and $\text{Tr}[ \hat{L}_{B}^2 \bar{\rho}(t)]$, corresponding to the QFI $\tilde{F}_{Q}^{(A)}$ and $\tilde{F}_{Q}^{(B)}$ of the two SLD contributions $ \hat{L}_{A}$ and $ \hat{L}_{B}$ defined above. Eqs.\,(\ref{LA})-(\ref{FQB}) are the main results of the paper. They provide a recipe to obtain the ultimate precision bound and optimal POVM that saturates it for {\em any} initial state in the set $\mathcal{K}_{\bf{\hat{z}}}$ and signal direction $\bf{\hat{m}}$. The precision scalings derived by Datta {\em et al} \cite{DattaPRA} in the absence of dephasing may be recovered from Eq.\,(\ref{FQB}) by computing the QCRB after setting $\kappa(t)=0$. Importantly, for $\theta \neq \pi/2$, 
one can readily check that  $\tilde{F}_{Q}^{(A)}$ gives the leading-order contribution to the QFI for all members of $\mathcal{K}_{\bf{\hat{z}}}$, and $\tilde{F}_{Q}^{(B)}[\rho(t)]$ may be neglected in the $J \gg 1$ limit we consider. In this case, the QFI takes the expression
$$ F_{Q}[\bar{\rho}(t)] 
\approx  2 \delta \,J^3\, t\, T \,\frac{\sin(2 \theta)^2 }{Q(\theta, t)},
\quad \theta \ne \pi/2, $$ 
and the metrological usefulness of the state is then tied to the effective quadrature squeezing parameter $\delta$ given in Eq.\,\eqref{delta}. This is nothing else than the HP limit of implementing a \emph{two-axis countertwisting} at the input \cite{Kita1993}, with the effective displacement $\eta$ not playing a role. Note that in the absence of an initial rotation, $\beta=0$, the state is entangled, but we would have $\delta=1$, just like for a classical CSS preparation. Rearranging of the quadratures through a rotation is thus crucial for precision enhancement. 

The achievable precision can be accessed by standard error propagation, Eq.\,(\ref{mom}), with $\mathcal{O}=J\hat{p}+ 2 J \eta \hat{x}$ the asymptotically optimal POVM. Physically, this may be thought of as a nonlinear readout of the form 
\begin{equation}
\mathcal{O}=e^{i \eta J_x^2/2} \,J_y \,e^{-i \eta J_x^2/2} \approx N\hat{p}/2 + N \eta \hat{x}, 
\label{readout}
\end{equation}
which is realized in practice by applying an anti-squeezing operation $e^{i \eta  \hat{x}^2}$ right before measurement. In the absence of temporal correlations, expanding  the resulting precision in powers of $\gamma t \ll 1$ shows that all states in $\mathcal{K}_{\bf{\hat{z}}}$ are metrologically equivalent, leading to an optimal scaling $\Delta \hat{b}_{\text{opt}} \propto J^{-1}$, as already found for a CSS and a $\Phi$ state. For non-Markovian dynamics, optimizing $\Delta \hat{b}(t)$ with respect to both the measurement time and the signal direction and re-expressing in terms of $N=2J$, we obtain 
\begin{equation}
\Delta \hat{b}^{\text{OATS}}_{\text{opt}} = (3 \sqrt{3}/4)^{1/2}  (\kappa_0 \omega_c/T)^{1/2}  \delta^{-1/4} N^{-5/4},
\label{eq:opt}
\end{equation}
which is achieved for 
\begin{eqnarray}
\tau_{\text{opt}} &=& \frac{1}{2 \kappa_0 \omega_c} |\csc (\theta_{\text{opt}}) | (N \delta/2)^{-1/2}  , \label{optpar} \\
\theta_{\text{opt}}&= &  \text{arccos}  (\sqrt{2/3}). 
\end{eqnarray}
Notably, the above equation reveals that the optimal angle $\theta_{\text{opt}}$ between the signal and the initial state {\em differs} from the noiseless case, where $\theta^0_{\text{opt}}= \pi/4$, but remains state-independent. Especially interesting is the case of a {\em perfect echo} (PE) protocol, corresponding to $\beta=-\pi/2$, $\mu = N^{-1/2}$, 
which was shown to reach optimal scaling for linear metrology in both a noiseless scenario \cite{Monika} and under collective dephasing \cite{third}; here, by using Eq.\,(\ref{mom}), we find that the interaction-based readout of Eq.\,\eqref{readout}, corresponding to $\eta =-N^{-3/2}$, leads to 
$$\Delta \hat{b}_{\text{opt}}^{\text{OATS,PE}} \propto N^{-3/2}, \quad\text{Zeno regime},$$ 
matching the scaling that can be achieved through the maximally entangled, non-Gaussian $\Phi$ state. Note that this state lies at the fringe of the HP regime of applicability, when the excitations $\langle a^{\dagger}a \rangle $ are of the same order of magnitude as $N$.

Optimal asymptotic precision bounds in terms of the total number of sensors, $N$, for some representative members of $\mathcal{K}_{\bf{\hat{z}}}$, as well for $\Phi$ states, are summarized in Table \ref{table:1}.

\section{Ratio estimator for quadratic metrology}
\label{ratio}

The figure of merit used throughout this work as a measure of metrological performance is the QCRB, which quantifies the ultimate precision one may achieve in estimating $b$ by measuring $\nu$ times a fixed POVM. Specifically, in the asymptotic $N \gg1$ regime we consider, the bound can be saturated by measuring an observable $\mathcal{O}$ which is diagonal in the eigenbasis of the SLD $\hat{L}$. Obtaining the precision in terms of the estimator $\hat{b}$ that is linked to the method of moments is generally problematic in the presence of noise, however, and improved estimators are needed. While no general solution to this problem is available to the best of our knowledge, for collective dephasing noise we show here that the strategy we introduced for frequency estimation in the linear setting \cite{FUR} can be generalized to nonlinear settings as well. In line with our previous analysis, we focus on the quadratic case, $k=2$.

\subsection{Maximally entangled initial state}

The basic ideas may be illustrated in their simplest form by considering an initial $|\Phi\rangle$ state. In a noiseless setting, the generalized HL of Eq.\,\eqref{quadrhl} is reached by measuring the survival probability $\mathcal{O}_{|\Phi \rangle}$. Since $\langle \mathcal{O}(\tau)\rangle = \cos(N^2 b \tau/4)= f(b)$, one can estimate the target frequency with the estimator 
 \begin{align*}
    \hat{b}= 4\,(\tau N^2)^{-1} \arccos(\langle \hat{\mathcal{O}}(\tau)\rangle) =f^{-1}(\hat{f}(b)).
 \end{align*}
However, when dephasing is present, the mean value is modified by a decay factor, $\langle \mathcal{O}(\tau)\rangle= e^{-N^2 \kappa(\tau)/4}\,\cos(N^2 b \tau/4)\equiv h(b) $. Disregarding the effect of the noise, by still letting $\hat{b}= f^{-1}(\hat{h}(b))$,
leads to a biased estimator. Alternatively, employing the correct inversion formula,
\begin{align*}
    \hat{b} & \myeqb 4(\tau N^2)^{-1} \text{arccos}\big( e^{J^2\kappa(\tau)}\,\hat{h}(b)\big) = h^{-1}(\hat{h}(b) ) ,
\end{align*}
would require {\em exact} knowledge of the decay coefficient $\kappa(\tau)$. Furthermore, even if that were possible, the construction of a bona-fide estimator remains problematic, as the argument of the arc-cosine may {\em a priori} yield imaginary values for certain detection outcomes \cite{FUR}, making the estimator ill-defined.

These issues can be circumvented by also measuring a second observable, $\mathcal{O}'\equiv \mathcal{O}_{|\Phi'\rangle} = |\Phi' \rangle \langle \Phi'|- (I-|\Phi' \rangle \langle \Phi'|)$, where $|\Phi'\rangle = (| J,J\rangle_{\bf{\hat{z}}} + i | J,0 \rangle_{\bf{\hat{z}}} )/\sqrt{2}$ is a different generalized GHZ state, the same number of times -- for a total of $2\nu$ measurements, over a runtime of $2T$. Since $\langle \mathcal{O'}(\tau)\rangle =  e^{-N^2 \kappa(\tau)/4} \sin(N^2 b \tau/4)$, we can solve for $b$ from the quotient of expectation values, which is independent upon the decay factor. This gives rise to the following \emph{ratio estimator} $\hat{b}_{\text{R}}$:
\begin{align*}
    \hat{b}_{\text{R}}^{\Phi} (\tau)\equiv 4 \, (N^2 \tau)^{-1} \arctan\Bigg( \frac{\langle \hat{\mathcal{O}'}(\tau)\rangle}{  \langle \hat{\mathcal{O}}(\tau)\rangle} \bigg).
\end{align*}
In the local estimation setting, the uncertainty may then be computed by error propagation in a way similar to Eq.\,\eqref{mom}, that is, $\Delta \hat{b}_{\text{R}}^2(\tau) =\nu^{-1} \sum_{i} \big[ \partial_{\langle u_i \rangle} \hat{b}_{\text{R}}\vert_{b_0}\big]^2 \Delta u_i^2 (\tau)$, 
with $u_1= \mathcal{O}(\tau ), u_2= \mathcal{O}'(\tau )$. The resulting uncertainty, 
\begin{align*}
    \Delta \hat{b}_{\text{R}}^{\Phi} (\tau)^2  &= \frac{16 e^{ N^2 \kappa(\tau)/2}+4 \cos( N^2 b_0 \tau)-1}{ N^4 T \tau},
\end{align*}
matches the QCRB, $\Delta \hat{b}_{\text{R}}^{\Phi}(\tau)^2= (\nu F_{\rm Q}[\bar{\rho}_{\rm \Phi}(t)] )^{-1}$, which we derived in Sec.\,\ref{noisyb}, at an optimal operating point where $b_0$ $\cos(N^2 b_0 \tau)=1$. It follows that the ultimate scaling bounds for $\Phi$ states discussed in Sec.\,\ref{noisyb} are enforced.

\subsection{Product initial state}
\label{ratioCSS}

\begin{figure*}[t!]
\centering
\includegraphics[width=18cm]{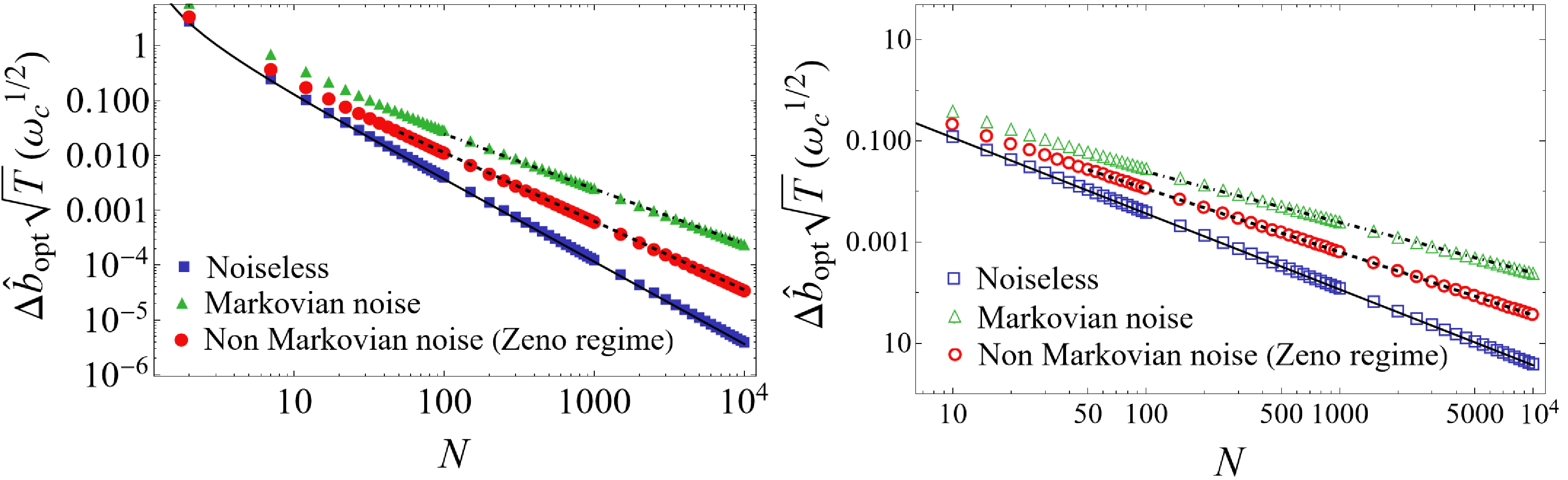}
 \vspace*{-5mm}
\caption{
{\bf Precision scaling of the QCRB vs.\,ratio estimator for an initial product state.} The precision of $|\text{CSS}\rangle_{\bf \hat{q}} = e^{i (\theta-\pi/2) J_y} |\text{CSS}\rangle_{\bf \hat{z}}$ as a function of particle number is numerically optimized over both $\theta$ and the encoding time derived from measuring $J_y$ (left) and the ratio estimator (right) for.  In each case, the black curves denote the analytically computed asymptotic performance. Specifically: noiseless dynamics, $\kappa(t)=0$ (blue squares), $\Delta \hat{b}_{\text{opt}}^{\text{CSS},0} \propto N^{-3/2}$ [Eq.\,\eqref{quadrsql}]; collective non-Markovian dephasing in the Zeno regime, $\kappa(t) \approx \kappa_0^2 (\omega_c t)^2$ (red circles), $\Delta \hat{b}_{\text{opt}}^{\text{CSS}} \propto N^{-5/4}$ [Eq.\,\eqref{dbCSSb}];
collective Markovian dephasing, $\kappa(t)= \gamma t$ (green triangles), $\Delta \hat{b}_{\text{opt}}^{\text{CSS}} \propto N^{-1}$ [Eq.\,\eqref{dbCSS}, green triangles].
In all cases, the ratio estimator achieves the same performance at the cost of doubling the total experimental runtime, $2T$. Here, $\kappa_0^2 = 3$, $\omega_c=1$, $\gamma=3$. 
}
\label{Ratio} 
\end{figure*}

Consider now an arbitrary initial state in $\mathcal{C}_{\bf{\hat{n}}}(\theta)$, namely, a CSS with variable polar angle and $\phi=0$. As we have already noted in Sec.\,\ref{noisyb}, exact formulas for the mean values $\langle J_x(t)^n\rangle$ and $\langle J_y(t)^n\rangle$, $n \in \{1,2\}$ can still be derived in the presence of collective noise. For example, we have 
\begin{align}
\langle J_x(t) \rangle &=  \frac{1}{4} \,N e^{-\kappa(t)} \sin (\theta ) \left( \zeta +\zeta^*\right), 
\label{Jxt} \\
\langle J_y(t) \rangle &=  \frac{1}{4i} N e^{-\kappa(t)} \sin (\theta )\left( \zeta- \zeta^*\right),  
\label{Jyt} 
\end{align}
where the quantity $\zeta$ is given by
\begin{align*}
\zeta&= \bigg[ e^{-i b\, t} \sin ^2\bigg(\frac{\theta }{2}\bigg)+ e^{i b\, t} \cos ^2\bigg(\frac{\theta }{2}\bigg)\bigg]^{N-1},
\end{align*}
and the combinations $\zeta \pm \zeta^*$ are purely real or imaginary quantities, respectively, expressible in terms of Chebyshev polynomials of the first and second kind, $T_n(\cos(\theta)) \equiv \cos(n \theta)$ and $\sin(\theta ) U_n( \cos(\theta))= \sin((n+1) \theta)$ [see Eqs.\,\eqref{zeta1}-\eqref{zeta2}].
Aside from issues due to noise-induced bias, we note that the estimator obtained by inverting Eq.\,(\ref{Jyt}) after $\nu$ measurements of the $J_y$ spin component, which saturates the QCRB for an optimal CSS with $\theta= \pi/4$, results in a rather unwieldy expression even in the noiseless case. 

In the noisy scenario, imagine we measure both $J_x$ and $J_y$, each a number $\nu \gg 1$ of times. Then, we may estimate $b$ from the ratio $\langle \hat{J}_y(\tau)\rangle/\langle \hat{J}_x(\tau)\rangle$ under the assumption that we can fix the operating point at $b_0=0$. In this way, we have $b=b_0+ \delta b=\delta b$, and Eqs.\,(\ref{Jxt})-(\ref{Jyt}) may be expanded in terms of the smallness parameter $\delta b \,\tau \ll 1$.
Using the fact that, in this limit, $(1+\cos(\theta)^2 \tan( \delta b \,\tau)^2)^{-1/2} \approx  \cos(\delta  b\,\tau \cos(\theta))$, we obtain
\begin{align*}
\frac{\langle J_y(\tau) \rangle}{\langle J_x(\tau) \rangle} & \approx 
\frac{ \sin(\delta b\;\tau \cos(\theta) ) U_{N-2}\big( \cos(\delta b\;\tau \cos(\theta) ) \big)}{T_{N-1}\big(\cos(\delta b\;\tau \cos(\theta) )\big)} \\
    &= \tan\big[ \delta b\;\tau (N-1) \cos(\theta)  \big].
\end{align*}
It follows that an asymptotically unbiased ratio estimator is given in this case by
\begin{align*}
\hat{b}_{\text{R}}^{\text{CSS}}(\tau)
\equiv \big[(N-1)\,\cos(\theta)\,\tau\big]^{-1} \arctan\left( \frac{\langle \hat{J}_y(\tau)\rangle}{\langle \hat{J}_x(\tau)\rangle}  \right).
\end{align*}
One may then compute the uncertainty linked to $ \hat{b}_{\text{R}}$ through error propagation as usual. This leads to 
\begin{align}
\Delta \hat{b}_{\text{R}}^{\text{CSS}}(\tau)^2 = \frac{\langle J_y(\tau)\rangle^2\, \Delta J_x^2(\tau) + \langle J_x(\tau)\rangle^2\, \Delta J_y^2(\tau) }{N^2 \, T \, \tau\, \cos(\theta)^2\, \left( \langle J_x(\tau)\rangle^2 +\langle J_y(\tau)\rangle^2  \right)^2}, 
\label{DeltabR}
\end{align}
where exact expressions for all terms entering $\Delta \hat{b}_{\rm R}^{\text{CSS}}(\tau)$ are given in Appendix\,\ref{app:CSS1}. The asymptotic behavior of Eq.\,(\ref{DeltabR}) can be then obtained numerically by optimizing over $\theta$ and $\tau$. As shown in Fig.\,\ref{Ratio}, the resulting performance of this ratio estimator matches that of the QCRB in the noiseless limit and in the presence of both Markovian and temporally correlated collective dephasing, also confirming the optimality of the operating point $b_0=0$. We further support these numerical results with analytical derivations in Appendix \ref{app:perf}.

Finally, we remark that extending this construction to entangled OATS states is rather tedious on account of the cumbersome computations involved, yet conceptually straightforward. Performing a second-order cumulant expansion over the qubit degrees of freedom in the asymptotic $N \gg 1$ limit, as done for the linear encoding case \cite{Bias}, an analytic approximation for $\langle J_x(\tau)\rangle ,\langle J_y(\tau)\rangle$ matching the exact mean values to great accuracy can be obtained. Derivation of a ratio estimator along similar lines to those described here then follows.

\section{Conclusion}
\label{conclusion}

We have quantified the impact of collective dephasing from a classical noise environment on frequency estimation using generalized Ramsey protocols with quadratic signal encoding. Our main findings may be summarized as follows. 

First, dephasing noise degrades the scaling of the achievable estimation precision to a degree dependent on the noise properties. In the Markovian limit, the absence of temporal correlations bounds all Gaussian squeezed states to perform at a $N^{-1}$ asymptotic scaling, irrespective of whether or not the initial state is entangled; sub-$N^{-1}$ scaling may be achieved in the presence of temporally correlated noise and non-Markovian dynamics in the short-time Zeno regime, even with unentangled input states and conventional Ramsey measurements. In this case, the use of a suitable OATS in conjunction with an anti-squeezing operation before readout yields the same optimal asymptotic $N^{-3/2}$ precision scaling that maximally entangled generalized GHZ states can afford. Altogether, these results support the conjecture that, under temporally correlated collective ``one-body'' dephasing ($p=1$ in the notation of Ref.\,\onlinecite{DelCampo}), quadratic signal encoding (more generally, nonlinearities of order $k\geq 2$) may offer a scaling advantage beyond the {\em linear} HL, but may still encounter a \emph{no-go for ``generalized superclassical'' precision} -- if the generalized classical bound is appropriately redefined as the one attainable with product states, $N^{-3/2}$ (more generally, $N^{-(k-p/2)}$, $k\geq 2$). While we do not pursue a formal proof here, we expect it may be obtained using ideas and techniques similar to those employed for the linear case \cite{third}.

Second, we have also called attention on the fact that collective dephasing introduces bias in the estimator that is linked to the measurement of a single system operator, within the standard local-estimation setting based on the methods of moments. Extending our linear-metrology approach \cite{Bias}, we have shown that an alternative, asymptotically unbiased {\em ratio estimator} reaching the same precision as the standard one may be constructed, provided we measure \emph{two} distinct observables and effectively double the measurement resources.  

Directly related to the above, it is worth noting that coupling to a \emph{nonclassical} environment -- for instance, bosonic modes as relevant in trapped-ion settings \cite{Bohnet} -- leads to an additional source of estimation bias. In this case, bath-mediated entanglement generation would be possible, resulting in an additional ``Lamb-shift'' term of the form $\dot{\Psi}(t) J_z^2$ in the main Hamiltonian [Eq.\,(\ref{Ham})] with $\Psi(t)$ being a time-dependent coefficient engendered by the anti-symmetrized noise spectrum \cite{multiDD,FelixPRA}. In turn, this would introduce an \emph{additive} bias that is absent in the linear metrology scenario, as both $b$ and $\dot{\Psi}(t)$ would couple to the system through the same operator. Devising a strategy capable of countering this additional source of bias is an interesting question for the future.  

More generally, an outstanding issue in nonlinear quantum metrology is how to exploit interactions among sensors to better estimate a parameter coupled to an {\em external} field, as opposed to some intrinsic properties of the sensing system itself. A recent proposal \cite{Generating} suggests measuring ancillary bosonic modes to generate an effective quadratic nonlinearity on the system, that could then be used to sense the rotation induced by an external field (see also Appendix \ref{app:imp}). Although the ensuing output states are not deterministic, which complicates a mapping to our model, it would be worth exploring the extent to which a similar scheme could be brought to bear on the noisy estimation problem we considered here.  It is our hope that our work will prompt additional investigation in the implementation of nonlinear sensing protocols and their potential in the presence of realistic noise sources.

\begin{acknowledgments}

Support from the U.S. National Science Foundation under Award No.\,PHY-2013974 and the U.S. Army Research Office through U.S. MURI Grant No.\,W911NF1810218 is gratefully acknowledged. Part of this research was performed while the authors were visiting the Institute for Mathematical and Statistical Innovation (IMSI), which is supported by the National Science Foundation (Grant No.\,DMS-1929348).
\end{acknowledgments}


\begin{appendix} 

\onecolumngrid

\section{Remarks on physical implementations}
\label{app:imp}

\subsection{The bosonic Josephson junction }
\label{app:BJJ}

BECs have been heavily studied as a system for implementing precision interferometry \cite{Choi,BoixoNonLinearBEC,SmerziRMP}. At sufficiently low temperature, the condensate is well described by the single-particle wavefunction $\psi({\bf{r}},t)$, which obeys the Gross-Pitaevskii equation,
\begin{eqnarray*}
     i \frac{\partial}{\partial t} \psi({\bf{r}},t)= \left[- \frac{1}{2m} \nabla^2 + V_{\text{ext}}({\bf r}) +\frac{4\pi  a_s}{m}  |\psi({\bf{r}},t)|^2 \right]\psi({\bf{r}},t), 
\end{eqnarray*}
where $V_{\text{ext}}({\bf{r}},t)$ is the external potential, $m$ the particles mass, and  $U= 4 \pi a_s/m$ the two-body interaction strength, that depends on the $s$-wave scattering length $a_s$.  In the Josephson approximation, the bosons are restricted to occupy a two-dimensional Hilbert space spanned by modes $|a\rangle$ and $|b\rangle$. The mechanism to generate the modes may be external, when they are spatially localized by confining the system in a double-well potential \cite{Khor2009}; or internal, when the condensate is realized in a single harmonic trap and the atoms can populate two different hyperfine states \cite{BoixoNonLinearBEC}. In this limit, 
the system is well described by the number-conserving {\em bosonic Josephson junction} (BJJ) Hamiltonian \cite{Khor2009,SmerziRMP,Leggett2001}:
\begin{eqnarray}
    H_{\text{BJJ}}= - \Omega J_x + \chi J_z^2 + \delta J_z. 
    \label{HBJJ}
\end{eqnarray} 
Here, $\delta$ represents an energetic imbalance between the modes, which can be arranged to be zero, 
whereas the effective interaction strength $\chi$ and the amplitude $\Omega$ are tunable control parameters
\cite{BECCollDeph,SmerziRMP}. Physically, in the external BJJ, $\Omega$ quantifies the degree of tunneling between the two wells, whereas in the internal BJJ it corresponds to a Rabi frequency between the two hyperfine states. In both cases, the interaction strength is proportional to a combination of scattering lengths: $\chi= (U_{aa}+U_{bb}- 2 \,U_{ab}) $ for the internal BJJ and $\chi=(U_{aa}+U_{bb})$ for the external BJJ \cite{Ferrini2011}, where $U_{ij}\equiv (4 \pi a_s^{(i,j)})/m$ and $a_s^{(i,j)}, i,j \in \{ a,b\}$, is the mode-dependent scattering length, with the cross term corresponding to interspecies scattering \cite{SmerziRMP}. Eq.\,(\ref{HBJJ}) is in direct correspondence with the two-mode description via a standard Jordan-Schwinger mapping, $J_x \mapsto (a^{\dagger} b - a b^{\dagger})$, $J_y \mapsto i (a b^{\dagger}- a^{\dagger} b)/2 $, and $J_z \mapsto  (a^{\dagger} a - b^{\dagger} b)/2 $, where $a^{\dagger} (b^{\dagger})$ creates a bosonic excitation in mode $|a\rangle$ ($|b\rangle$), respectively.

In realistic setups, phase randomness and dephasing noise arise due to variations of external control fields, instabilities in the trapping potential, and collective spin excitations \cite {Khor2009,Ferrini2011,BECCollDeph}. 
Tuning $\Omega=0$ in Eq.\,(\ref{HBJJ}) and assuming that the fluctuations may be modeled by a classical stochastic process $\xi(t)$, the dynamics are described by
$     H_{\rm S}= \chi J_z^2 + \xi(t) J_z ,$ which corresponds to the Hamiltonian in Eq.\,(\ref{Ham}) with $k=2$ and $b \equiv \chi$. Thus, estimation of the interaction strength $\chi$ may be carried out with precision surpassing the linear HL in principle.
  
\subsection{Towards nonlinear metrology with conditional operations}
\label{app:gen}

As stressed in the main text, it would be desirable to generate a quadratic sensing Hamiltonian which allows estimation of an external field parameter, as opposed to some intrinsic property of the sensors themselves. A recent proposal \cite{Generating} makes use of two bosonic modes, a probe $P$ and an ancilla $A$, to achieve this. The modes have creation, annihilation, number, position and momentum operators $\{ a_i^{\dagger}, a_i, \hat{n}_i ,\hat{x}_i, \hat{p}_i \}$ $i\in \{ A,P\}$, with $\hat{n}_i=a_i^{\dagger}a_i$. Initially, the ancilla is prepared in a squeezed vacuum state, with squeezing strength $r$, $e^{-i r (a_A^2-a_A{\dagger \, 2})/2}|0\rangle_A$, after which a rotation $R(\theta)= e^{i \theta \hat{n}_A}$ is applied. The ancilla is then entangled with a probe state $|\psi \rangle_{P}$ through the gate $C_R\equiv e^{-i g \hat{p}_A \otimes \hat{n}_P}$. Measurement of the position operator $\hat{x}_A$ yielding outcome $m$ leads then to a probe post-measured state which can be written as a commuting product of four operators acting on $|\psi\rangle_P$: an outcome-dependent rotation $U_c(m, \theta)\equiv e^{-i g m  \cot(\theta) \hat{n}_P/2}$, a deterministic {\em Kerr-like nonlinear interaction} $U(\theta)\equiv e^{i g^2 \cot(\theta) \hat{n}_P^2/4} $, and two non-unitary terms which are vanishing in the limit of sufficiently high squeezing, $r \gg 1$. Finally, implementing a corrective rotation $U_c^{\dagger}(m,\theta)$, the effective evolution is described exclusively by an quadratic Hamiltonian which can be used for estimating an angle $\theta$. When a probe with mean Fock number $\bar{n}_P$ is prepared in a coherent state $|\alpha\rangle_P$, the authors predict a precision $\Delta \hat{\theta} \propto \bar{n}_P^{-3/2}$ in the absence of noise.

The above protocol may be realized by a  quantum light-matter interface where a two-mode polarized light beam passes through an atomic ensemble in the HP limit \cite{Generating}. In such a scenario, the gate $C_R$ can be generated by a Faraday interaction, $H_F= J_z F_Y,$ which couples the $\bf{\hat{y}}$ component of the light Stokes vector to the $\bf{\hat{z}}$ component of the atomic angular momentum, whereas, crucially, the rotation $R(\theta)$ is induced by a uniform {\em external} magnetic field $B_0 \,\bf{\hat{z}}$ via a Zeeman interaction, $H_Z= \vec{B} \cdot \vec{J}$. The angle $\theta \equiv b \tau$ then represents the accumulated phase acquired by the probes, which precess at the Larmor frequency $b= |B_0 \,\gamma|$, with $\gamma$ being the gyromagnetic ratio, after evolving under $H_Z$ for a finite encoding time $\tau$.

Importantly, the above scheme relies on measuring ancillary degrees of freedom, which induces a statistical ensemble of conditional probe final states. This complicates the analysis of the protocols performance, as the expression we have obtained for the QFI by assuming a deterministic final state must be modified. 
While performing a complete study is beyond our present scope, the scheme described above may be adapted to the estimation of a target frequency $b$, as we focused on. An analysis of the impact of collective dephasing may be carried out by considering random fluctuations of the magnetic field during the encoding, $H_Z(t)= [b+ \xi(t)] J_z$, and by exploiting tools from local quantum estimation of a post-measurement ensemble to obtain the QFI.

\section{Noiseless precision bounds for product vs.\,maximally entangled quantum states}
\label{app:noiseless}

\subsection{CSS evolution under quadratic dynamics}
\label{app:CSS1}

Given the Hamiltonian in Eq.\,\eqref{Ham} in the main text with $k=2$, we consider the set $\mathcal{C}_{\bf{\hat{n}}}(\theta,\phi)$ of all CSS.  While our first aim is to re-derive known precision bounds for the noiseless setting, for later use we show how to obtain expectation values of relevant observables in the case where dephasing noise is also present. The density operator at a generic time $t$ may be written as  
\begin{align}
    \bar{\rho}_{\text{CSS}}(t)&=\sum_{m,m'=-J}^J \rho_{m,m'}(0)\,e^{-i b\,t\,  (m^2-m'^2)} e^{-\kappa(t) (m-m')^2} |J, m\rangle_{\bf{\hat{z}}}\, _{\bf{\hat{z}}} \langle J, m'|, 
\label{rhoCSS}
\end{align}
where the decay factor $\kappa(t)$ depends on the noise spectral properties as in Eq.\,\eqref{freqchi} and the matrix elements of the initial state in the Dicke basis read \cite{Nori2}:
\begin{align}
 \rho_{m,m'}(0)=   \bigg[1+ \tan\bigg (\frac{\theta}{2}\bigg)^2\bigg]^{-2J}\! \bigg[ e^{i \phi} \tan \bigg(\frac{\theta}{2} \bigg) \bigg]^{J+m} \!\bigg[ e^{-i \phi} \tan \bigg(\frac{\theta}{2} \bigg)  \bigg]^{J+m'} \sqrt{ {{2J}\choose{J+m}} {{2J}\choose{J+m'}}}.
 \end{align}
Below we evaluate the first and second moments of the spin components orthogonal to the signal direction, $\langle J_i^k(t)\rangle$, $i \in \{x,y\}$, $k \in \{1,2\}$, which are instrumental in computing the best performance that can be achieved by interferometry protocols with classical input states. Breaking down $J_x= \frac{1}{2} (J^+ + J^-),\, J_y=\frac{1}{2 i} (J^+-J^-) $, with $J^{\pm}$ being the usual raising (lowering) operators, and assessing each component individually \cite{Kita1993}, the first moments are found to be given by the expressions quoted in the main text, Eqs.\,\eqref{Jxt}-\eqref{Jyt}, where, for general non-zero $\phi$, the quantity $\zeta$ is given by
\begin{align*}
     \zeta&=  e^{-i \phi } \bigg[ e^{-i b\, t} \sin ^2 \bigg(\frac{\theta }{2}\bigg)+e^{i  b \,t} \cos ^2\left(\frac{\theta }{2}\right)\bigg]^{2J-1}.
\end{align*}
The second moments may also be evaluated in a similar fashion,
\begin{align}
\langle J^2_x(t) \rangle &= \frac{1}{8} \bigg[ 
J \big((1-2 J) \cos (2 \theta )+2 J+3\big) + 
J (2 J-1) e^{-4 \kappa(t)} \sin ^2(\theta ) \left( \Theta+\Theta^* \right)\bigg],
\label{JxSq2}\\
\langle J^2_y(t) \rangle &= \frac{1}{8} \bigg[
J \big( (1-2 J) \cos (2 \theta )+2 J+3\big) - 
J (2 J-1) e^{-4 \kappa(t)} \sin ^2(\theta ) \left( \Theta+\Theta^* \right)\bigg], 
\label{JySq2}\\
\:\Theta & \equiv  e^{-2 i \phi } \bigg[ e^{-2 i \, b \,t} \sin ^2\left(\frac{\theta }{2}\right)+e^{2 i \,b\, t} \cos ^2\left(\frac{\theta }{2}\right)\bigg]^{2 J-2}. \nonumber
\end{align}

It is worth pointing out that closed analytic expressions showing that $ \zeta \pm \zeta^*,  \Theta+\Theta^*$ (and hence $\langle J_x(t)^n\rangle$ and $\langle J_y(t)^n\rangle$, $n \in \{1,2\}$), are {\em real-valued} quantities, may be derived. The key idea consists on rewriting the relevant expressions in polar form. For instance, we have $e^{-i \phi} \left[ e^{-i \,b\, t} \sin ^2\left(\frac{\theta }{2}\right)+e^{i\, b\, t} \cos ^2\left(\frac{\theta }{2}\right) \right] \equiv r e^{i \alpha}$ for some suitable $r \geq0, \alpha$; then, $\zeta=  e^{-i \phi } r^{2J-1} e^{i \alpha (2J-1)} $. This expression may be broken down into real and imaginary parts, and then evaluated in terms of Chebyshev polynomials of the first and second kind, which obey $T_n(\cos(x))= \cos(n x)$ and $U_n(\cos(x)) \sin(x)= \sin((n+1)x)$, respectively. We get:
\begin{align}
\zeta + \zeta^* &= \left[\cos( b\, t)^2+\cos(\theta)^2 \sin(b\, t)^2 \right]^{J-1/2} \left[ \cos (\phi )\, T_{2J-1} \left( u\right)+ u\, \cos (\theta ) \sin (\phi ) \tan (b\, t)\; U_{2J-2}(u)  \right], 
\label{zeta1} 
\\
\zeta - \zeta^* &= \left[\cos(b\, t)^2+\cos(\theta)^2 \sin( b\, t)^2 \right]^{J-1/2} \left[ \sin (\phi )\, T_{2J-1} \left( u\right)- u\, \cos (\theta ) \cos (\phi ) \tan (b\, t)\; U_{2J-2}(u)  \right], 
\label{zeta2}
\\
\Theta + \Theta^* &= \left[\cos(2 \, b\, t)^2+\cos(\theta)^2 \sin(2 \, b\, t)^2 \right]^{J-1}\,\left[\cos (2 \phi ) T_{2J-2}\left( u' \right)+  u' \sin (2 \phi ) \cos (\theta ) \tan (2 \, b\, t) \,  U_{2J-3}\left( u' \right) \right], 
\label{Theta} 
\end{align}
where the arguments $u,u'$ are respectively given by $u\equiv (\cos ^2(\theta ) \tan ^2(\delta b\; t)+1)^{-1/2}$ and $u'\equiv  \left(\cos ^2(\theta ) \tan ^2(2  \delta b\; t)+1\right)^{-1/2}$.

\subsection{Noiseless CSS asymptotic precision}
\label{app:CSS2}

Let us now show assume that $\xi(t)\equiv 0$ and show that the generalized SQL scaling for quadratic encoding, $\Delta \hat{b} \propto J^{-3/2}$, can be saturated by a member of the set $\mathcal{C}_{\bf{\hat{n}}}(\theta,\phi)$. 
We can compute the QFI of any CSS $|\theta, \phi\rangle$ as $F_Q[\rho_{\text{CSS}}(\tau)]= 4 \nu (\langle J^4_z(\tau)\rangle- \langle J_z^2(\tau) \rangle^2 ) = 4 \nu \Delta J_z^2(\tau)$. The mean values $\langle J^2_z(\tau)\rangle$, $\langle J^4_z(\tau)\rangle$ in the limit of no detuning, $\delta b=0$, are easily found to be
\begin{align}
\langle J^2_z(\tau)\rangle &=\frac{J}{4} \big[1+2J +(2J-1) \cos(2\theta)\big], \label{jz1} \\
    \langle J^4_z(\tau)\rangle &= \frac{J}{32} \left[ J-1+ 12\, J^2 (J+1)+4 (2J-1)^2 (J+1) \cos(2\theta) +(J-1)(2J-3)(2J-1) \cos(4 \theta) \right], 
\label{jz2}
\end{align}
by 
using Eq.\,(\ref{rhoCSS}) with $\kappa(\tau)=0$. This leads to
\begin{eqnarray}
F_Q[\rho_{\text{CSS}}(\tau)]= \frac{T \,\tau \,J}{2}\, (2J -1) \sin(\theta) \, \big[4J-1 + (4J -3) \cos(2 \theta)\big]. 
\label{QFICSS}
\end{eqnarray}
Naturally, the result is $\phi$-independent due to the Hamiltonian's azimuthal symmetry. The optimal polar angle maximizing Eq.\,(\ref{QFICSS}) for any finite $J$ may be obtained exactly,
\begin{eqnarray*}
\theta_{\text{opt}}(J) = \pm \arctan \bigg(\sqrt{\frac{2J-1}{2J-2}}\bigg)\, \underset{J\rightarrow \infty}{\longrightarrow} \,\theta_{\text{opt}}= \pm \frac{\pi}{4},
\end{eqnarray*} 
leading to the optimal direction ${\bf{\hat{n}}}_{\text{opt}}= \sin(\theta_{\text{opt}}) {\bf{\hat{x}}}+ \cos(\theta_{\text{opt}}) {\bf{\hat{z}}}  \approx \tfrac{1}{\sqrt{2}} (\bf{\hat{x}}+\bf{\hat{z}} )$.
Replacing into Eq.\,(\ref{QFICSS}) leads, via the QCRB, to the following precision bound:
\begin{eqnarray}
\Delta \hat{b}(\tau) \geq \sqrt{\frac{4 J -3}{T \tau J (2J -1)^3}} \simeq \frac{1}{\sqrt{2\,T \tau\, J^3 }}, \qquad J\gg 1,
\label{QCRBCSS}
\end{eqnarray}
in agreement with Eq.\,\eqref{quadrsql} in the main text. A setup realizing the bound in the asymptotic regime for $|\text{CSS}\rangle_{\bf{\hat{n}}_{\text{opt}}}$ consists of measuring $J_y$ at readout. The uncertainty can then be evaluated by error propagation, using Eq.\,\eqref{mom}; specifically, 
\begin{eqnarray}
\Delta \hat{b}(\tau) = \frac{\sqrt{\langle J^2_y (\tau) \rangle - \langle J_y (\tau) \rangle^2 }}{ \sqrt{\nu} \, 
|\partial_{b} \langle J_y (\tau) \rangle \vert_{b_0} |}. 
\label{Ramsey}
\end{eqnarray}
By using the explicit expressions of the mean values in Eqs.\,(\ref{Jxt}), \eqref{Jyt} and (\ref{JySq2}) for $\kappa(\tau)=0$ and working at at operating point $b_0=0$, we obtain:
\begin{eqnarray*}
\Delta \hat{b}(\tau) = \frac{\sqrt{2}}{\sqrt{T \tau}  \sqrt{J}\, (2J -1)}.
\end{eqnarray*}
Note how this is lower bounded by the QCRB, Eq.\,(\ref{QCRBCSS}), to which it rapidly converges for increasing $J$.

\subsection{Noiseless asymptotic precision of $\Phi$ states}
\label{ent}

If we consider the generalized GHZ state $\Phi$ defined in the main text, the corresponding QFI may be easily obtained in terms of the variance of $J_z^2$, $F_Q[\rho_{\Phi}(\tau)]= 4 \nu \, \Delta (J_z^2)^2(\tau) = 
4\nu\big[ \langle J_z^4(\tau)\rangle- \langle J_z^2(\tau)\rangle^2 \big]= \nu J^4$, and leads, via the QCRB, to the following bound to the precision:
\begin{eqnarray}
\Delta \hat{b} (\tau) \geq \frac{1}{\sqrt{T \tau} \, J^{2}}. 
\label{Phinsless} 
 \end{eqnarray}
This coincides with Eq.\,\eqref{quadrhl}, defining the generalized HL for quadratic encoding. Note that, like the corresponding expression for the CSS obtained above,  Eq.\,(\ref{Phinsless}) is monotonically decreasing in $\tau$ and thus does not provide an optimal time to measure. The precision is instead maximized by prolonging the encoding period. In practice, technical limitations impose an upper bound to the detection time.  

For this state, the QCRB may be saturated by implementing a measurement of the survival probability. That is, we measure the observable $\mathcal{O}= |\Phi\rangle \langle \Phi | - (I-|\Phi\rangle \langle \Phi |) $. The relevant mean values and uncertainty may be easily computed to be $ \langle \mathcal{O}(t) \rangle = \cos(J^2 b t)$ and $\langle \mathcal{O}^2(t)\rangle = 1$, implying a generalized HL, as stated: 
\begin{align}
\Delta \hat{b}(\tau) = \frac{\Delta \mathcal{O}(t)}{\sqrt{\nu}|\partial_b \langle \mathcal{O}(\tau) \rangle \vert_{b_0}}  = \frac{\sqrt{ \sin(J^2 b_0 \,\tau)^2}}{ \sqrt{T \tau} J^2 | \sin(J^2 b_0 \,\tau) |}= \frac{1}{\sqrt{T \tau} J^2},
\end{align}
independently of the operating point $b_0$ around which we measure.

\section{Noisy asymptotic precision scaling for product vs.\,maximally entangled quantum states}
\label{app:noisyphi}

\subsection{Noisy CSS asymptotic precision}
\label{app:noisyCSS}

The performance of the noiseless optimal strategy using CSS states may be also exactly assessed in the presence of collective dephasing, $\kappa(\tau) \neq 0$.  Evaluating  uncertainty Eq.\,(\ref{Ramsey}) for $\theta=\pi/4$ and again working at $b_0=0$,
we find:
\begin{eqnarray}
    \Delta \hat{b}(\tau)=\frac{2 J \sinh(2\kappa(\tau))+ e^{2 \kappa(\tau)}+ \cosh(2 \kappa(\tau))}{T\, \tau \,J^3}. 
    \label{deltabgen}
\end{eqnarray}
Importantly, the presence of decay factor $\kappa(\tau)$ defines a $J$-dependent optimal measurement time, $\tau_{\text{opt}}^{\text{CSS}}$, at which the uncertainty $\Delta \hat{b}_{\text{opt}}^{\text{CSS}} \equiv \Delta \hat{b}(\tau_{\text{opt}}^{\text{CSS}})$ reaches a minimum. The noise spectral properties, which enter through $\kappa(t)$, heavily impact the ultimate asymptotic scaling of precision that may be reached, as we show below. 

\smallskip

\begin{figure*}[t!]
\centering
\includegraphics[width=18cm]{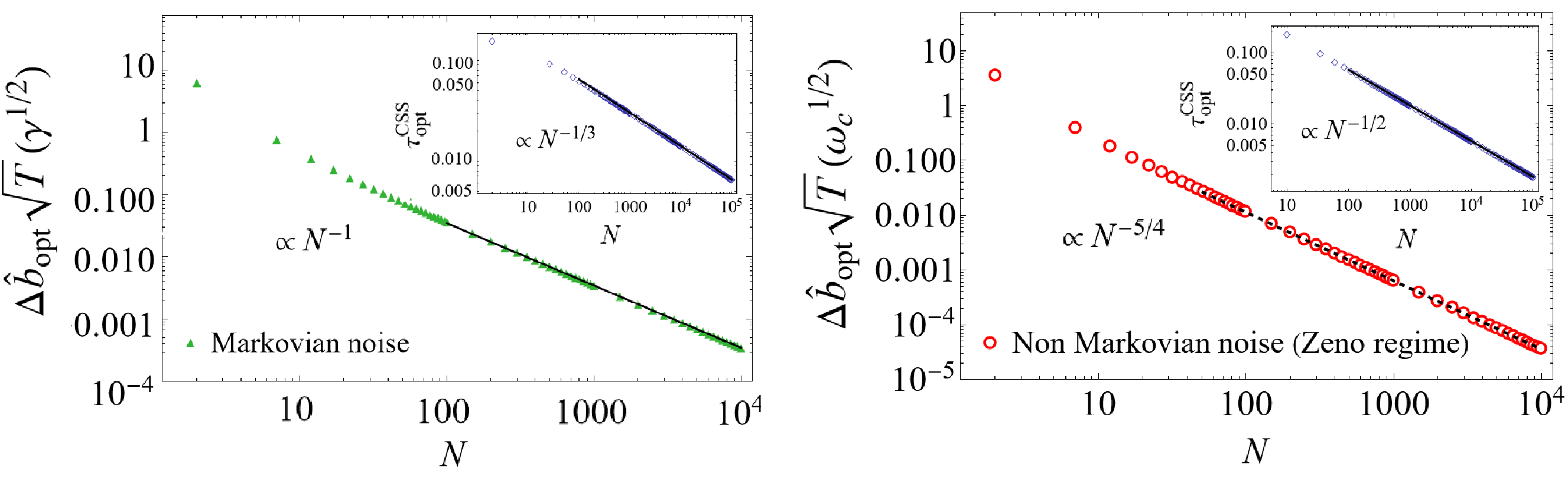}
 \vspace*{-8mm}
\caption{{\bf Performance of optimal noiseless protocol under collective dephasing.} Time-optimized uncertainty for an initial CSS $|\frac{\pi}{4},0\rangle$ and a $J_y$-measurement as a function of sensors' number. This selection of initial state and measurement basis saturates the QCRB, $\Delta \hat{b}_{\text{opt}}^{\text{CSS}} \propto N^{-2}$ in the large-$N$ limit, in the absence of decoherence. For Markovian dephasing (left, green triangles), the performance is degraded to $\Delta \hat{b}_{\text{opt}}^{\text{CSS}} \propto N^{-1}$. For non-Markovian dephasing dynamics in the Zeno dynamics (right, hollow red circles), we have $\Delta \hat{b}_{\text{opt}}^{\text{CSS}} \propto N^{-5/4}$. The insets show the optimal encoding time as a function of $N$. Parameters: 
$\kappa_0^2 = 3$, $\omega_c=1$, $\gamma=3$. }
\label{noiseless} 
\end{figure*}

\paragraph{Markovian noise.}
In the absence of temporal correlations, $\kappa(t)= \gamma t$. Taking the time derivative of Eq.\,(\ref{deltabgen}) and equating it to zero to solve for $\tau^{\text{CSS}}_{\text{opt}}$, one obtains the condition
\begin{align*}
    2J (x \cosh(x)-\sinh(x))+(x \sinh(x)-\cosh(x)-e^{x} (1-x)) \approx 4 \left(\frac{J}{3} x^{3}-1 \right)=0, \quad  x=2 \gamma t, 
\end{align*}
where the second approximate equality comes from expanding in the limit of $x \ll 1, J x^3 \sim 1$. Solving for the optimal measurement time yields,
\begin{eqnarray*}
    \tau^{\text{CSS}}_{\text{opt}}= \frac{3^{1/3}}{2}\,\gamma^{-1} \,J^{-1/3} 
\end{eqnarray*}
Replacing this expression into Eq.\,(\ref{deltabgen}), and expanding to leading order in $J \gg 1$, we find:
\begin{align}
  \Delta \hat{b}_{\text{opt}}^{\text{CSS}} = \sqrt{\frac{\gamma}{T}}\, J^{-1}. 
  \label{dbCSSM}
\end{align}
It follows that scaling is degraded to the linear HL, $\Delta \hat{b} \propto J^{-1}$, a $J^{-1/2}$ loss with respect to the noiseless limit.
Numerical optimization confirms both of these scalings to be correct, as shown in Figure \ref{noiseless} (left).

\smallskip

\paragraph{Zeno dynamics.}
For non-Markovian noise in the Zeno regime, $\omega_c t \ll 1$ such that $\kappa(t) \simeq \kappa_0^2 (\omega_c t)^2$, we may solve for $\tau_{\text{opt}}^{\text{CSS}}$  and the resulting optimal uncertainty $\Delta \hat{b}^{\text{CSS}}_{\text{opt}}$ by expanding the numerator and denominator in Eq.\,(\ref{deltabgen}) with respect to time to second and first order, respectively. One finds:
\begin{eqnarray*}
\Delta \hat{b}(\tau) \approx  \frac{\sqrt{{J}/{2} + J^2 \kappa_0^2 (\omega_c \tau)^2}}{\sqrt{T \tau} J^2 }.
\end{eqnarray*}
Taking the time derivative of this approximate expression and equating it to zero, then evaluating $\Delta \hat{b}(\tau^{\text{CSS}}_{\text{opt}})$ in the asymptotic limit $J \gg 1$, we obtain:
\begin{align}
 \tau^{\text{CSS}}_{\text{opt}}= \frac{1}{\sqrt{2}} \frac{1}{\kappa_0 \omega_c}\,J^{-1/2} ,\qquad \Delta \hat{b}_{\text{opt}}^{\text{CSS}} = 2^{1/4} \sqrt{\frac{\kappa_0 \omega_c}{T}} J^{-5/4}, 
 \label{zeno}
\end{align}
Note that the different constant prefactor when comparing with $\Delta \hat{b}_{\text{opt}}^{\text{CSS}}$  in Table \ref{table:1} stems from the fact that before we allowed for an optimization over the angle between CSS and signal directions, whereas here we simply evaluated precision for the optimal \emph{noiseless} initial state where $\theta= \pi/4$ remains fixed. Our asymptotic expressions for $\tau^{\text{CSS}}_{\text{opt}}$ and $\Delta \hat{b}_{\text{opt}}^{\text{CSS}} $ are confirmed by optimizing $\Delta \hat{b}(\tau)$ numerically over the encoding time, see Fig.\,\ref{noiseless} (right).

\subsection{Noisy asymptotic precision of $\Phi$ states} 

Despite the fact that the generalized GHZ state $|\Phi\rangle$ is entangled, its time-evolved density operator remains effectively one of a two-level system in the presence of collective dephasing; explicitly, 
 \begin{eqnarray*}
 \bar{\rho}_{\Phi}(t) = \frac{1}{2} \left( |J,J \rangle_{\bf{\hat{z}}}\, _{\bf{\hat{z}}} \langle J,J| +e^{i  J^2 b t} e^{- J^2 \kappa(t)} |J,J \rangle_{\bf{\hat{z}}}\, _{\bf{\hat{z}}} \langle J,0| + e^{-i  J^2 b t } e^{- J^2 \kappa(t)} |J,0 \rangle_{\bf{\hat{z}}}\, _{\bf{\hat{z}}} \langle J,J|+|J,0 \rangle_{\bf{\hat{z}}}\, _{\bf{\hat{z}}}\langle J,0| \right).
 \end{eqnarray*}
It is thus possible to diagonalize it exactly, $\bar{\rho}_{\Phi}(t)= \sum_i p_i (t)|\lambda_i\rangle \langle \lambda_i|$, after which the SLD, $\hat{L}$, may be readily obtained:
\begin{equation*}
\hat{L} =2 \sum_{j,k} \frac{1}{ p_j (t) + p_k(t)} \langle \lambda_j| \partial_b\, \bar{\rho}_{\Phi}(t) | \lambda_k \rangle = i J t e^{-J^2 \kappa(t)} 
\big(|J,J\rangle_{\bf{\hat{z}}}\, _{\bf{\hat{z}}}\langle J, 0| -|J,0\rangle_{\bf{\hat{z}}}\, _{\bf{\hat{z}}} \langle J, J|\big) .
\end{equation*}
This leads to the QFI and, via the QCRB, to the desired precision bound
\begin{align*}
    F_Q[\bar{\rho}_{\Phi}(t)]= \nu\,\text{Tr}[\hat{L}^2  \bar{\rho}_{\Phi}(t)] = T t\, e^{-2 J^2 \kappa (t)} J^4 \quad \Rightarrow \quad \Delta \hat{b}(\tau) \geq \frac{e^{ J^2 \kappa(\tau)}} {\sqrt{T \tau}\, J^2},
\end{align*}
which is still saturated by a measurement of the survival probability. Computing the relevant mean values we find
$\langle \mathcal{O}(t)\rangle = e^{-J^2 \kappa(t)} \cos(J^2 b t)$ and $\langle \mathcal{O}^2(t)\rangle =1$,
leading to the uncertainty
\begin{align}
    \Delta \hat{b}(\tau)^2= \frac{1- e^{-2 J^2 \kappa(\tau)} \cos(J^2 b_0 \tau)} {T \tau e^{-2 J^2 \kappa(\tau)} \sin(J^2 b_0 \tau)}\;\;\;\; &\myeq \;\;\;\; \frac{e^{ 2 J^2 \kappa(\tau)}} {T \tau J^4}, 
    \label{deltaO}
\end{align}
where in the third equality we evaluated $\Delta \hat{b}^2(\tau)$ at the optimal operating point $\varphi (\tau)=J^2 b_0 \tau = \pi/2$.
To proceed further, we need to minimize this expression with respect to time. Once again, Markovian and temporally correlated noise exhibit different behavior.

\smallskip

\paragraph{Markovian noise.}
For Markovian noise, $\kappa(t)= \gamma t $, let us take the derivative with respect to time of the (squared) QCRB,
\begin{align*}
   \frac{d}{dt}\Delta \hat{b}(t)^2 = \frac{e^{2J^2 \gamma t}}{J^4 t T} (2 J^2 \gamma\,t-1) .
\end{align*}
Equating to zero yields the optimal measurement time $\tau_{\text{opt}}^{\Phi}=(2 \gamma J^2)^{-1}$. Finally, replacing in the QCRB we find:
\begin{align}
\Delta \hat{b}_{\text{opt}}^{\Phi}=\sqrt{\frac{\gamma}{T}}\,\sqrt{2\,e}\, J^{-1}. 
\label{Dbphi}
\end{align}
Thus, the deleterious effects of Markovian dephasing worsen the scaling of the optimal noiseless strategy with entangled states from $J^{-2}$ to $J^{-1}$. The latter is the same asymptotic performance CSS achieve under this type of decoherence; moreover, the constant prefactor in Eq.\,(\ref{Dbphi}) is larger than the one in Eq.\,(\ref{dbCSSM}), meaning that an entangled initial preparation is actually detrimental for precision in this scenario.

\smallskip

\paragraph{Zeno dynamics.}
For temporally correlated noise in the Zeno limit $\kappa(t) \approx \kappa_0^2 (\omega_c t)^2$, equating the squared QCRB time derivative to zero yields 
\begin{align*}
  \frac{d}{dt}\Delta \hat{b}(t)^2 = \frac{e^{ 2 J^2 \kappa_0^2 (\omega_c t)^2}}{J^4 T t}\, \big[4 J^2 \kappa_0^2\, (\omega_c t)^2 -1\big]=0.
\end{align*}
This determines the optimal time $\tau_{\text{opt}}^{\Phi}=\frac{1}{2}\,(\kappa_0\,\omega_c\, J)^{-1}$, which in turns provides the asymptotically best performance
\begin{eqnarray*}
\Delta \hat{b}^{\Phi}_{\text{opt}} = \sqrt{2} e^{1/4}  \sqrt{\frac{\kappa_0 \omega_c}{T}}\, J^{-3/2}.
\end{eqnarray*}
This $J^{-3/2}$ power law is in-between the linear and quadratic HLs. Thus, temporally correlated dephasing is less harmful than its Markovian counterpart for entangled states as well.

\section{Derivation of asymptotic QFI for arbitrary Gaussian states}
\label{app:}

\subsection{Equivalence between fixed Hamiltonian and fixed initial state settings}
\label{equivalence}

In our original statement of the problem, we are interested in quantifying the sensing performance of a certain class of spin squeezed states, to be specified below, evolving under the action of a noisy quadratic Hamiltonian with a \emph{fixed} quantization axis. Here, for convenience, we take said axis to be along the $\bf{\hat{x}}$ direction,
\begin{align}
    \tilde{H}(t)=b\, J_x^2+ J_x \xi(t) \equiv \tilde{H}_{\bf \hat{x}}(t).
    \label{Ham3}
\end{align}
Clearly, the Hamiltonian in Eq.\,(\ref{Ham}) in the main text can be recovered by performing a $\pi/2$ rotation about the $\bf{\hat{y}}$ axis, $H(t)= e^{i \pi/2 J_y } \tilde{H}(t) e^{-i \pi/2 J_y } $, hence the metrological equivalence between the dynamics each of them give rise to is evident. 

Let us consider the squeezed state of Eq.\,\eqref{oatz}, $|\text{OATS}\rangle_{\bf{\hat{z}}} \equiv e^{-i \beta J_z} e^{-i \mu J_x^2}  |\text{CSS}\rangle_{\bf{\hat{z}}}$ with associated density matrix $\rho_{\bf{\hat{z}}}(\beta, \mu)$, where we say the OATS is along $\bf{\hat{z}}$ as the squeezing and 
rotation operations are performed to a CSS in the north pole of the generalized Bloch sphere. The class $\mathcal{G}_{\bf{\hat{q}}}(\theta) $ of initial preparations we are interested in can be generated by performing a rotation $\mathcal{R}(\theta) \equiv e^{i (\theta-\pi/2) J_y}$  on $|\text{OATS}\rangle_{\bf{\hat{z}}}$:
\begin{align}
 |\text{OATS}\rangle_{\bf{\hat{q}}} \equiv   e^{i (\theta- \pi/2) J_y} |\text{OATS}\rangle_{\bf{\hat{z}}}  = e^{-i \beta J_{\bf{\hat{q}}}}  e^{-i \mu J^2_{\bf{\hat{p}}}} |\rm{CSS}\rangle_{\bf{\hat{q}}},  \quad 
 \bf{\hat{q}}= \cos(\theta) \bf{\hat{x}} + \sin(\theta) \bf{\hat{z}}, \; \bf{\hat{p}}= \sin(\theta) \bf{\hat{x}} - \cos(\theta) \bf{\hat{z}} .
\end{align}
Crucially, this transformation preserves the parameters $\beta, \mu$ of the original state acting,  however, over a CSS rotated along $\bf{\hat{n}} \neq \bf{\hat{z}}$. We may then write:
$$\mathcal{G}_{\bf{\hat{q}}}(\theta) \equiv \big\{ \rho_{\bf{\hat{q}}}(\mu, \beta, \theta): \rho_{\bf{\hat{q}}}(\mu, \beta,\theta) =\!\mathcal{R}(\theta)\rho_{\bf{\hat{z}}}(\mu, \beta) \mathcal{R}^{\dagger}(\theta)\big \}.$$
It is then  easy to show that the expectation value of an observable $\mathcal{O}$ with respect to a state in $\mathcal{G}_{\bf{\hat{q}}}(\theta)$ evolving under Hamiltonian $\tilde{H}_{\bf \hat{x}}(t)$ in Eq.\,(\ref{Ham3}) [scenario A, fixed-axis Hamiltonian] can be equivalently written as the expectation value of a rotated observable $\tilde{O}(\theta) \equiv  \mathcal{R}^{\dagger}(\theta) \mathcal{O} \mathcal{R}(\theta)$ with respect to an
OAT of identical parameters $\beta, \mu$ but \emph{fixed} direction $\bf{\hat{z}}$,  $|\text{OATS} \rangle_{\bf{\hat{z}}} \in \mathcal{G}_{\bf{\hat{z}}}$, evolving under a Hamiltonian with a variable quantization axis $\tilde{H}_{\bf{\hat{m}}}(t)= b\,J_{\bf{\hat{m}}}^2+\xi(t)\,J_{\bf{\hat{m}}}$ [scenario B, fixed-axis initial state]. Explicitly: 
\begin{align}
    \langle \mathcal{O} (t) \rangle &\equiv \big \langle \,_{\bf{\hat{q}}}\langle \text{OATS}| e^{i \tilde{H}_{\bf \hat{x}}(t)}\, \mathcal{O}\, e^{-i \tilde{H}_{\bf \hat{x}}(t)} |\text{OATS} \rangle_{\bf{\hat{q}}} \big\rangle_{\xi} \nonumber\\
    &= \big \langle \,_{\bf{\hat{z}}}\langle \text{OATS}| \,(\mathcal{R}^{\dagger}(\theta) e^{i \tilde{H}_{\bf \hat{x}}(t)} \mathcal{R}(\theta))\,[  \mathcal{R}^{\dagger}(\theta) \mathcal{O}  \mathcal{R}(\theta)]\,  (\mathcal{R}^{\dagger}(\theta) e^{-i \tilde{H}_{\bf \hat{x}}(t)} \mathcal{R}(\theta))\,|\text{OATS} \rangle_{\bf{\hat{z}}} \big\rangle_{\xi}\nonumber\\
    &=\big\langle \,_{\bf{\hat{z}}}\langle \text{OATS}|  e^{i \tilde{H}_{\bf{\hat{m}}}(t)}\, [\tilde{\mathcal{O}}(\theta) ]\, e^{-i \tilde{H}_{\bf{\hat{m}}}(t)} |\text{OATS} \rangle_{\bf{\hat{z}}} \big\rangle_{\xi}, 
    \quad 
    \bf{\hat{m}}= \sin(\theta)\bf{\hat{x}}+\cos(\theta) \bf{\hat{z}}. 
%
\end{align} 

We now show that, in the light of this equivalence, the ultimate sensitivity bounds that can be attained in scenario A and scenario B are the same. 
Let us first assume that we compute the optimal performance of a state $|\rm OATS \rangle_{\bf{\hat{q}}} \in \mathcal{G}_{\bf{\hat{q}}}(\theta)$, evolving under the Hamiltonian $\tilde{H}_{\bf \hat{x}}(t)$. The QCRB can be accessed by minimizing the CRB over all possible POVMs, and is saturated by measuring an observable ${L}$ which is diagonal in the SLD eigenbasis. In the local regime we consider, the resulting precision $\Delta \hat{b}(\tau)$ can be expressed in terms of expectation values $\langle {L}(\tau)\rangle$, $\langle {L}^2(\tau)\rangle$ through Eq.\,(\ref{mom}). Alternatively, we may evaluate the optimal uncertainty of an initial $|\rm OATS \rangle_{\bf{\hat{z}}}$ evolving under the Hamiltonian $H_{\bf{\hat{m}}}(t)$. Exploiting the equivalence between scenario A and scenario B, it follows that by measuring $\tilde{L}(\theta) \equiv  \mathcal{R}^{\dagger}(\theta) {L} \mathcal{R}(\theta)$, the same $\Delta \hat{b}(\tau)$ value may be reached here, as claimed. 
Writing $\tilde{H}_{\bf{\hat{m}}}(t)$ in the HP limit, which forces us to restrict the initial states to the subset $\mathcal{K}_{\hat{\bf{z}}}\subset \mathcal{G}_{\hat{\bf{z}}}$ of \emph{properly squeezed} states, we obtain the Hamiltonian in Eq.\,(\ref{HHP}) in the main text, which we have used as the starting point of our analysis.

\subsection{Quantum mechanics in phase space}
\label{app:phasespace}

The phase-space formulation provides an equivalent description of quantum mechanics, alternative to the usual one relying on operators acting on a Hilbert space. Here we present a brief summary of its most salient features, as useful in our context. We focus on a one-dimensional continuous system $\rm{S}$ with associated Hilbert space $\mathcal{H}_{\rm S}$, which corresponds to the HP limit carried in the main text. However, generalizations to higher dimensions and discrete variables are well-known in the literature \cite{Agarwal,Bizarro,Hillery}.

The correspondence between both formulations is given by the Weyl transform \cite{pedestrians}, which maps an operator $\mathcal{A}$ acting on $\mathcal{H}_{\rm S}$ into the phase-space function $f_{\mathcal{A}}(x,p)$, which depends on the canonical position and momentum coordinates $x$ and $p$:
\begin{eqnarray}
f_{\mathcal{A}}(x,p)  = \frac{1}{\pi} \int_{-\infty}^{\infty}\, dy \, \langle x + y | \,\mathcal{A}\, | x-y \rangle\, e^{-2 i p y}= \frac{1}{\pi} \int_{-\infty}^{\infty}\, du \, \langle p + u | \,\mathcal{A}\, | p-u \rangle\, e^{2 i u x}, 
\label{Weyl}
\end{eqnarray}
where $\mathcal{A}$ can be written in the position (momentum) operator eigenbasis -- right hand-side of the first (second) equality in (\ref{Weyl}). Importantly,  polynomial functions of position and momentum operators transform straightforwardly: $f_{\mathcal{A}}(x,p)= (a x+ b p)^n$, for $\mathcal{A}= (a \,\hat{x}+ b\, \hat{p})^n$, $a,b \in \mathbb{C}, n \in \mathbb{N}$. It can also be shown that the trace of two operators $\mathcal{A}, \mathcal{B}$ is equivalent to the phase-space integral of the product of their corresponding Weyl transforms:
\begin{eqnarray}
\text{Tr}[\mathcal{A} \mathcal{B}]= \int \!  dx \!\int \! dp \, f_{\mathcal{A}}(x,p)\,f_{\mathcal{B}}(x,p).  
\label{intphsp}
\end{eqnarray}
 As noted in the main text, the Wigner function is defined as the Weyl transform of the 
 density operator.
It provides a \emph{pseudo-probability} distribution, in the sense that it is real-valued and normalized to unity but may attain \emph{negative} values across pockets of phase space; negativity of the Wigner function is indeed often used as an indicator of non-classicality. It follows from Eq.\,(\ref{intphsp}) that the expectation value of any operator $\mathcal{A}$ can be expressed in terms of an average over phase space of its Weyl transform, weighted by the Wigner function:
\begin{eqnarray*}
\langle \mathcal{A} (t) \rangle= \int \!  dx \!\int \! dp \, f_{\mathcal{A}}(x,p)\,W_{\bar{\rho}}(x,p,t). 
\end{eqnarray*}
The prescription that assigns the Weyl transform of a product of operators, say, ${\mathcal{D}= \mathcal{A} \mathcal{B}}$,
in terms of their individual Weyl transforms, is set by the \emph{Moyal product} \cite{pedestrians}, 
\begin{eqnarray}
f_{\mathcal{A} \mathcal{B}} (x,p)& \equiv f_{\mathcal{A}}(x,p) \,\star\, f_{\mathcal{B}}(x,p) =f_{\mathcal{A}}(x,p) \,\exp\left( \frac{i}{2} \Lambda \right)\, f_{\mathcal{B}}(x,p)= \sum_{n \geq 0} \frac{1}{n!} \left(\frac{i}{2}\right)^n f_{\mathcal{A}}(x,p) \, \Lambda^n \, f_{\mathcal{B}}(x,p)  , 
\label{Moyal}
\end{eqnarray}
where the operator $\Lambda$, acting on functions $f_{\mathcal{A}}(x,p)$ and $f_{\mathcal{B}}(x,p) $, is defined by
\begin{eqnarray*}
    f_{\mathcal{A}}(x,p) \, \Lambda \, f_{\mathcal{B}}(x,p) \equiv \partial_x  f_{\mathcal{A}}(x,p)  \partial_p f_{\mathcal{B}}(x,p) - \partial_p  f_{\mathcal{A}}(x,p)  \partial_x f_{\mathcal{B}}(x,p)  .
\end{eqnarray*}

Thanks to the above formalism, we may transform the operator equality that implicitly defines the SLD operator in the Hilbert space representation, Eq.\,\eqref{equality}, into a standard differential equation and ultimately solve for the asymptotic precision limits (Sec.\,\ref{app:ultimate}). It is also possible to an equation of motion that directly governs the evolution of the Wigner function behavior in time. For the system with quadratic encoding and collective dephasing we consider, whose state evolves under the master equation given in Eq.\,\eqref{ME}, 
one finds that $W_{\bar{\rho}}(x,p,t)$ obeys a Fokker-Planck-like equation:
\begin{eqnarray}
\frac{\partial }{\partial t} W_{\bar{\rho}}(x,p ,t) =\big[b \sin(\theta) \big(2 \sin(\theta)\, \hat{x} - J  \cos(\theta)\big)\partial_p   + \dot{\kappa}(t) \, \partial_p^2\,\big]\, W_{\bar{\rho}}(x,p,t). 
\label{WeqofM}
\end{eqnarray}
This equation, together with the initial condition $W(x,p,0)= f(x,p)$, fully characterizes the system's evolution. The effect of collective dephasing as inducing diffusion along the 
momentum direction in this limit is particularly obvious from Eq.\,(\ref{WeqofM}).

\subsection{Computation of the Wigner function}
\label{app:compWig}

As a starting point, let us write the Wigner function  of the CSS in the HP approximation. Since, as noted in the main text, $|\text{CSS}\rangle_{\bf{\hat{z}}}$ is the ground state of $H=a^{\dagger} a$ in the low excitation limit, we can write state $|0\rangle$ in the position representation as  
\begin{eqnarray*}
|\text{CSS}\rangle_{\bf{\hat{z}}} \mapsto |0 \rangle\equiv  \left({J \pi}\right)^{-1/4}\! \int_{-\infty}^{\infty}\!dx \, e^{-x^2/(2J)} \,  |x\rangle  .
\end{eqnarray*} 
 Transforming into phase space, we find
 \begin{eqnarray*}
 W_{\text{CSS}}(x,p,0)= {\pi}^{-1}\,e^{-x^2/J- J p^2}.
\end{eqnarray*} 
For the rotated OATS given by $|\text{OATS} \rangle_{\bf{\hat{z}}} = 
e^{-i \beta (J-a^{\dagger} a)} e^{-i \mu \hat{x}^2} |0\rangle $, computation of the initial state in the position representation becomes more involved:
\begin{eqnarray}
    |\text{OATS} \rangle_{\bf{\hat{z}}} &= & e^{-i \beta (J-a^{\dagger} a)} e^{-i \mu \hat{x}^2} \, \left({J \pi}\right)^{-1/4} 
    \int_{-\infty}^{\infty}\!dx \, e^{-x^2/(2J)} \, 
     |x\rangle = e^{-i \beta (J-a^{\dagger} a)} \left({J \pi}\right)^{-1/4} \int_{-\infty}^{\infty}\!dx \, e^{-x^2(1/(2J)+ i \mu)}  |x\rangle \nonumber \\
    & =& \left({J \pi}\right)^{-1/4}  \,e^{-i \beta J} \int_{-\infty}^{\infty}\!du |u \rangle 
    \int_{-\infty}^{\infty}\!dx  \,e^{-x^2(1/(2J)+ i \mu)} 
    \sum_{n \geq 0} e^{i \beta n} \langle u |n \rangle \langle n| x \rangle ,
\label{mid}
\end{eqnarray}
where in the second line we have inserted two resolutions of the identity, in the position and number basis, for convenience. Using the explicit form of the number-states wavefunctions, in terms of the $n$th order Hermite polynomial $H_n(x)$, 
\begin{eqnarray*}
    \langle  x|n \rangle = \left( {\pi J} \right)^{-1/4} \frac{1}{\sqrt{2^n n!}} e^{-x^2/(2J)}\, H_n(x/\sqrt{J}),
\end{eqnarray*}
we find 
\begin{equation}
  |\text{OATS} \rangle_{\bf{\hat{z}}} =\left({J \pi}\right)^{-1/4} e^{-i \beta J} \int_{-\infty}^{\infty}\!du |u \rangle 
  \int_{-\infty}^{\infty}\!dx \, e^{-x^2(1/(2J)+ i \mu)} \,
  \sum_{n \geq 0 } \frac{e^{i \beta n}}{\sqrt{\pi J}}  \frac{1}{2^n n!} e^{-(u^2+x^2)/(2J)} H_n(x/\sqrt{J}) H_n(u/ \sqrt{J})  .
  \label{mid2}
  \end{equation}
By letting $r\equiv e^{i \beta}$, we can then use the \emph{Mehler's kernel} \cite{Durkin},
\begin{equation*}
   \sum_{n \geq 0} \frac{1}{n!}\,(r/2)^n\,H_n(x) H_n(y) e^{-(x^2+ y^2)/2} = \frac{1}{\sqrt{1-r^2}} \exp \bigg[ \frac{4 x y r - (1+r^2) (x^2+y^2)}{2 (1-r^2)} \bigg] , 
\end{equation*}
to evaluate the series. Replacing in Eq.\,(\ref{mid2}), and computing the resulting Gaussian integral, we find
 \begin{eqnarray}
   |\text{OATS} \rangle_{\bf{\hat{z}}} &=&\left({J \pi}\right)^{-1/4}  \,e^{-i \beta J} \int_{-\infty}^{\infty}\!du |u \rangle 
    \int_{-\infty}^{\infty}\!dx \, e^{-x^2(1/(2J)+ i \mu)} 
    \frac{1}{\sqrt{1-e^{2 i \beta}}} \exp\bigg[ \frac{4 u x - (1+e^{2 i \beta}) (u^2 + x^2)}{2 J (1-e^{2 i \beta})} \bigg]  
    \nonumber\\
    &= &\left({J \pi}\right)^{-1/4}  \,e^{-i \beta J} \int_{-\infty}^{\infty}\!du |u \rangle \,
    \frac{1}{\sqrt{1+2 e^{i \beta} J \mu \sin(\beta)}} \exp\bigg[ u^2 \frac{-i + J \mu (1+e^{2 i \beta}) }{2 J (i + J \mu (e^{2 i \beta}-1))}\bigg] .
 \end{eqnarray}
Having an expression for $|\text{OATS} \rangle_{\bf{\hat{z}}}$ in the position eigenbasis, it is now straightforward to perform the Weyl transform, 
and evaluate the Wigner function. One finds:
\begin{eqnarray}
& W_{\text{OATS}}(x,p,0) = {\pi}^{-1}  e^{- \frac{1}{J \delta} x^2 + J \delta (p+ 2 \eta x + )^2   }. 
\label{CIW}
\end{eqnarray}
Accordingly, we may regard the exponent as the Hamiltonian of a classical harmonic oscillator $H(x, p')$ of unit frequency and mass $m\equiv J \delta$, where the parameters $\delta$ and $\eta$, the latter entering the pseudo-momentum coordinate $p'\equiv p+ 2 \eta x$, both depend on the OATS rotation and squeezing angles $\beta, \mu$, in the form of Eqs.\,\eqref{eta}-\eqref{delta} given in the main text.

The initial condition in Eq.\,(\ref{CIW}), together with the equation of motion in Eq.\,(\ref{WeqofM}) fully determine the Wigner function at posterior times. We can solve for $W_{\text{OATS}}(x,p,t)$ using a standard Green's function approach, where we consider the function $G(x,p,t)\equiv X(x,p,t) P(x,p,t)$, which satisfies Eq.\,(\ref{WeqofM}) together with an ``impulse'' initial conditions: $X(x,p,0)= \delta (x)$ and $ P(x,p,0)= \delta(p)$. As there are no derivatives with respect to position on the Wigner equation of motion, we may simply write 
\begin{eqnarray}
 X(x,p,t) = \lim_{\sigma \to 0} \frac{1}{\sqrt{2 \sigma^2 \pi}} \, e^{- x^2/ (2 \sigma^2)}.
\end{eqnarray} 
For the same reason, the equation for $P(x,p,t)$ reduces to Eq.\,(\ref{WeqofM}), and can be easily solved to give
\begin{eqnarray}
P(x,p,t)= \frac{1}{2 \sqrt{\pi \kappa(t)}} \,e^{ - \frac{1}{4 \kappa(t) }  [ p- b t \sin(\theta) (2 \sin(\theta) x - J \cos(\theta)]^2 }.
\end{eqnarray}
The desired time-evolved Wigner function at time $t$, subject to initial condition Eq.\,(\ref{CIW}), may then be found by integrating
\begin{align}
W_{\text{OATS}}(x,p,t) &= \iint_{\mathbb{R}^2} \,G(x-x_0, p-p_0,t)\, W_{\text{OATS}}(x,p,0) \,dx_0\,dp_0 
\nonumber \\
&=  
{\frac{1}{\sqrt{\pi^2 Q(\theta, t)}}} \,e^{  -  \frac{1}{J \delta} x^2 + \frac{J \delta}{Q(\theta, t)} 
\big[ p+ 2 \eta x + 2\, \varphi\,\sin(\theta) \big( J \cos(\theta)+ x \sin(\theta) \big) \big]^2   }, 
\label{WignerG}
\end{align}
with $\varphi= b \,t$ and  $Q(\theta, t)$ given in Eq.\,\eqref{Qt} in the main text.

We stress that initial states that can be adequately described in this picture must obey the low-excitation condition $\langle a^{\dagger} a \rangle \ll J$, under which the HP representation is valid. The excitations number may be computed as 
\begin{eqnarray}
\langle a^{\dagger} a \rangle =\frac{1}{2}\,\iint W_{\text{OATS}}(x,p,t) \Big(\,\frac{1}{J} x^2 + J p^2 \Big) dx\, dp = \frac{1}{4} \left[ \delta^{-1}+ \delta (1+4 J^2 \eta^2) + 4 J \kappa(t) \right] \ll J. 
\label{exc}
\end{eqnarray}
Thus, the low-excitation requirement translates into restrictions on the allowed values of the squeezing angle $\mu$ and, therefore, $\delta, \eta$, as well as on the timescale over which our analysis is valid. By letting $\mu \equiv \mu_0 J^{-\alpha}$, with $\mu_0 \in \mathbb{R},\alpha >0$ in Eq.\,(\ref{exc}), we can see that the initial condition $t=0$ is already scaling like $J$ for $\beta = \pi/2, \alpha =1/2$: In particular, for $\mu = (2J)^{-1/2} $ as used in Ref.\,\onlinecite{Monika}, we find $\langle a^{\dagger} a \rangle \simeq J/2 $, at the limit of the range of applicability. It follows that $\alpha \leq 1/2$ for the HP regime to be valid, and the allowed initial states are properly squeezed: in other words, they do not wrap around the Bloch sphere, consistent with the fact that we are effectively approximating the dynamics to occur in the plane tangent to $\bf{\hat{z}}$.

\subsection{Ultimate precision bounds}
\label{app:ultimate}

We are now in a position to compute the SLD, and consequently derive the precision limits achievable by members of the set of properly squeezed Gaussian states in the HP limit. We focus on sensing small phase space displacements, $\varphi= \delta b \tau \ll 1$, at the operating point $b_0=0$. The first step is to transform the operator equality implicitly defining $\hat{L}$ in Eq.\,\eqref{equality}, that is, $2 \,\partial_b \bar{\rho}(t) = \hat{L} \bar{\rho}(t) + \bar{\rho}(t) \hat{L}$, into the phase-space language:
\begin{eqnarray}
-2 i t \sin(\theta) \left( \sin(\theta) f_{[\hat{x}^2, \bar{\rho}]}(x,p,t) + 2 J  \cos(\theta) f_{[\hat{x}, \bar{\rho}]}(x,p,t) \right) =  f_{\{\bar{\rho}, \hat{L}\}}(x,p,t) ,\label{SLDeq}
\end{eqnarray}
where the Weyl transforms of commutators and anti-commutators can be written in terms of the transform of the corresponding individual operators by means of Eq.\,(\ref{Moyal}), e.g, 
\begin{eqnarray*}
     f_{\{\bar{\rho}, \hat{L}\}}(x,p,t) = W_{\text{OATS}}(x,p,t) \star f_{L}(x,p,t) + f_{L}(x,p,t) \star W_{\text{OATS}}(x,p,t).
\end{eqnarray*}
Our goal is to solve for the Weyl transform of the SLD, $ f_{\hat{L}}(x,p,t)$.  Using the Moyal product, we can evaluate $ f_{[\hat{x}, \bar{\rho}]}(x,p,t)$ and $f_{[\hat{x}^2, \bar{\rho}]}(x,p,t)$ as follows:
\begin{eqnarray*}
f_{[\hat{x}, \bar{\rho}]}(x,p,t) = x \star W_{\text{OATS}}(x,p,t)- W_{\text{OATS}}(x,p,t) \star x = -i \,\partial_p W_{\text{OATS}}(x,p,t),    \\
     f_{[\hat{x}^2, \bar{\rho}]}(x,p,t) = x^2 \star W_{\text{OATS}}(x,p,t)- W_{\text{OATS}}(x,p,t) \star x^2 = - 2 i\, x \partial_p W_{\text{OATS}}(x,p,t).    
\end{eqnarray*}
So, the left hand-side of Eq.\,(\ref{SLDeq}) reads:
\begin{eqnarray}
 2 \,f_{\partial_b \bar{\rho}}(x,p,t) &=& 4 t  \sin(\theta) \big[ x \sin(\theta) + J \cos(\theta)  \big] \partial_p   W_{\text{OATS}}(x,p,t)  \nonumber \\
 &=& - 8 t\, J \delta \,
Q(\theta, t)^{-1} \, \sin(\theta) \,(p+ 2 \eta x) \, \sin(\theta) \big[ x \sin(\theta) + J \cos(\theta)  \big]    W_{\text{OATS}}(x,p,t), 
 \label{SLDLHS}
\end{eqnarray}
where we have exploited the Gaussianity of Eq.\,(\ref{WignerG}) to express $\partial_p  W_{\text{OATS}}(x,p,t)$ in terms of the Wigner function. To solve for $f_{\hat{L}}(x,p,t)$, as mentioned in the main text we pose the SLD $\hat{L}$ to be a linear combination of Hermitian operators at most quadratic in $\hat{x}$ and $\hat{p}$, $\hat{L} =  \sum_{q,j} \alpha^j_{q}\, q^{j}+ \beta \,\hat{x} \hat{p} \equiv a \hat{x} + b \hat{p} + c \{ \hat{x}, \hat{p}\} + e \hat{x}^2 + f \hat{p}^2,$ with $a,b,c,e,f \in {\mathbb R}$ undetermined coefficients. Transforming into phase space and  computing $f_{\{\bar{\rho}, \hat{L}\}}(x,p,t)$ using the Moyal product yields:
\begin{eqnarray*}
   W_{\text{OATS}}(x,p,t) \star f_{\hat{L}}(x,p,t) + f_{\hat{L}}(x,p,t) \star W_{\text{OATS}}(x,p,t)  = 2 (a x +b p + c x p + e x^2 + f p^2)+ \frac{1}{2} \left( c \,\partial_p \partial_x  - e\, \partial_p^2 - f \,\partial_x^2 \right)  W_{\text{OATS}}(x,p,t).
\end{eqnarray*}
Once again, Gaussianity makes the second order partial derivatives  proportional to $W_{\text{OATS}}(x,p,t)$. Equating the resulting expression to Eq.\,(\ref{SLDLHS}) and solving for $a, b,c,e$, we finally get
\begin{eqnarray}
f_{\hat{L}}(x,p,t)= -4 \,t  \sin(\theta) \bigg[ \frac{\delta  J^2  \cos(\theta) }{Q(\theta, t)} 
(p+ 2 \eta x) + 
\frac{\delta  J \sin(\theta)}{Q(\theta, t) +1}
\, \bigg( \frac{1}{2} p x + 2 \eta x^2 )\bigg) \bigg] . 
\end{eqnarray}
Transforming back into an operator acting on the systems Hilbert space, we may write $\hat{L}=-4 t \sin(\theta)(\hat{L}_A + \hat{L}_B)$, with:
\begin{eqnarray}
\hat{L}= -4 \,t  \sin(\theta) \left[ \frac{\delta  J^2 \cos(\theta)}{Q(\theta, t)} 
(\hat{p}+ 2 \eta \hat{x}) +  \frac{\delta  J \sin(\theta) }{
Q(\theta,t)+1} \bigg(\frac{1}{2} \{\hat{x}, \hat{p}\} + 2 \eta \hat{x}^2 \bigg) \right],
\label{L}
\end{eqnarray}
in agreement with Eqs.\,(\ref{LA})-(\ref{LB}) in the main text. Straightforward computation of $\text{Tr}[\bar{\rho}(t) \hat{L}^2]$ by rewriting $\hat{L}^2$ in phase space, through transformation rules stated in Appendix \ref{app:phasespace} then yields
\begin{eqnarray}
    F_Q[\bar{\rho}(t)]= 2\delta J^2 T t \sin(\theta)^2 \, 
    \left[ \frac{4 J \cos(\theta )^2}{4 \delta  J \text{$\kappa(t)$}\sin(\theta)^2+1} + 
    \frac{ \delta \sin ^2(\theta )}{2 \delta  J \text{$\kappa(t)$} \sin(\theta)^2+1} \right]
    = 2\delta J^2 T t \sin(\theta)^2  \left(  F_Q^{(A)}[\bar{\rho}(t)] +  F_Q^{(B)}[\bar{\rho}(t)]  \right) ,
    \label{FQ}
\end{eqnarray}
and, through the QCRB $\Delta \hat{b}(t) \geq (F_Q[\bar{\rho}(t)])^{-1/2}$ , the ultimate precision bounds for any member of $\mathcal{K}_{\bf{z}}$. 

Clearly, the scaling with $J$ of the precision $\Delta \hat{b}(t)$ depends on the rotation and squeezing angles of the initial OATS, $\beta$ and $\mu$, through quantities $\eta, \delta$ as defined in Eqs.\,(\ref{eta})-\eqref{delta}, as well as the measurement time $\tau$ and the angle $\theta$ between the signal direction and the $\bf{\hat{z}}$ axis. Direct substitution of Eqs.\,(\ref{eta})-\eqref{delta} into Eq.\,(\ref{FQ}) shows that $F_Q^{(A)}[\rho(t)]$ gives the leading-order contribution to the total QFI as it has a higher $J$-scaling than $F_Q^{(B)}[\rho(t)]$  for the parameter set $\theta \neq \pi/2$ and $\mu \propto J^{-\alpha}$, $\alpha < 1/2$, we are interested in (note that both terms scale equally for $\alpha=1/2$). It is then legitimate to optimize the $F_Q^{(A)}[\rho(t)]$ with respect to time and signal direction  while discarding the term $F_Q^{(B)}[\rho(t)]$ without detriment to overall asymptotic $J$-scaling: $\Delta \hat{b}(t)^2 \gtrsim F^{(A)}_{Q}[\rho(t)]^{-1}$. We stress, however, that even for $\alpha=1/2$, including both terms in the optimization only results in an slightly improved $J$-independent pre-factor for the overall precision, but does not otherwise change the main conclusions. For this reason, as it greatly complicates an analytical treatment, moving forward we do not include $F_Q^{(B)}[\rho(t)]$.

Assume that we consider temporally correlated noise in the Zeno regime, whereby the decay factor is quadratic in time. We can then show that
\begin{eqnarray*}
   \frac{d}{d t} (F_Q^{(A)}[\rho(t)])^{-1}=\frac{d}{d t} \delta J^2 T\, t \,\sin(\theta)^2 \, \frac{4 J \cos(\theta )^2}{4 \eta  J \text{$\kappa_0^2 (\omega_c t)^2$}\sin(\theta)^2+1}= 0 \quad \Rightarrow \quad
   \tau_{\text{opt}}= \frac{1}{2}  (\kappa_0 \omega_c)^{-1} |\csc (\theta )| (\delta  J)^{-1/2} ,
\end{eqnarray*}
gives the state-dependent optimal measurement time. In turn, optimizing $F_Q^{(A)}[\rho(\tau_{\text{opt}})] $ with respect to the signal angle yields
\begin{eqnarray*}
\frac{d}{d \theta} (2 \delta  J^3 T) \frac{1}{\omega_c} \cos(\theta )^2 \sin^2(\theta ) (\delta J \kappa_0^2)^{-1/2} = 0  \quad \Rightarrow \quad \theta_{\text{opt}}= \{ \pm\,  \,\text{arccos}(\pm \sqrt{2/3}) \}, 
\end{eqnarray*}
as stated in the main text. Replacing $\theta_{\text{opt}}$ in $F_Q^{(A)}[\rho(\tau_{\text{opt}})]^{-1} $, we find the optimal OATS precision given in Eq.\,\eqref{eq:opt}.
As Eq.\,\eqref{eq:opt} makes it clear, in the Zeno regime the dependence upon the initial state enters through the parameter $\delta$, given in Eq.\,\eqref{delta}; the parameter $\eta $, given in Eq.\,\eqref{eta},  is instead important in determining the optimal POVM. We recover the values given in Table I as follows. 

(1) The simplest case is $|\text{CSS}\rangle_{\bf{\hat{z}}}$, where $\beta_{\text{CSS}}=\mu_{\text{CSS}}=0$. Replacing in Eqs.\,\eqref{eta}, \eqref{delta}, and (\ref{eq:opt}), we find: 
\begin{eqnarray*}
\delta_{\text{CSS}}=1, \quad    \eta_{\text{CSS}}=0, \quad  \Delta \hat{b}_{\text{opt}}^{\text{CSS}} = \bigg(\frac{3\sqrt{3}}{4}\bigg)^{1/2}\,   \,\bigg(\frac{\kappa_0 \,\omega_c}{T}\bigg)^{1/2}  J^{-5/4}. 
\end{eqnarray*}

(2) For an OATS with minimal initial dispersion along the $\bf{\hat{y}}$-axis\cite{Kita1993}, we have $\mu_{\text{KU}}= 12^{1/6}  J^{-2/3}$, $\beta_{\text{KU}}= \pi/2- 3^{-1/6}2^{-1/3} J^{-1/3}$, leading to 
\begin{eqnarray*}
   \delta_{\text{KU}}=4\,2^{2/3}\, 3^{1/3}\, J^{2/3}, \quad    \eta_{\text{KU}}=\frac{ 1}{ 2^{7/3}\,3^{1/6}}\,J^{-4/3}, \quad \Delta \hat{b}^{\text{KU}}_{\text{opt}} = \bigg(\frac{3}{8}\bigg)^{1/2} \bigg(\frac{3}{2}\bigg)^{1/6}\, \left( \frac{\kappa_0 \omega_c}{T}\right)^{1/2} J^{-17/12} .
\end{eqnarray*}   

(3) For the initial states used in the PE protocol proposed in Ref.\,\onlinecite{Monika}, where $\mu_{\text{PE}}= (2J)^{-1/2} , \beta_{\text{PE}} = \pi/2$, which results in the following effective parameters $\delta, \eta$, and optimal precision: 
\begin{eqnarray}
   \delta_{\text{PE}}=2J, \quad    \eta_{\text{PE}}=- (2J)^{-3/2}, \quad \Delta \hat{b}^{\text{PE}}_{\text{opt}}= \frac{3}{4} \bigg(\frac{3}{2} \bigg)^{1/4}  \bigg(\frac{\kappa_0 \omega_c}{T} \bigg)^{1/2}  J^{-3/2} .
\end{eqnarray}

\section{Performance comparison between QCRB and ratio estimator}
\label{app:perf}

While the ratio estimator we introduced provides a simple, asymptotically unbiased estimator, an important question is to assess how the associated uncertainty $\Delta \hat{b}_{\text{R}}(\tau)$ compares against the QCRB, derived from the standard procedure of measuring the single observable $J_y$. Below we determine the ultimate bounds that may be reached with both schemes when sensing with CSS $|\theta, 0 \rangle$ in the presence of collective dephasing, after optimizing with respect to the encoding time and the polar angle.

\subsection{QCRB performance}

Our starting point is the uncertainty in Eq.\,(\ref{Ramsey}), with the relevant expectation values, Eqs.\,(\ref{Jxt})-(\ref{Jyt}) and Eqs.\,(\ref{JxSq2})-(\ref{JySq2}), evaluated for $\phi=0$ as well as $b_0=0$:
\begin{align}
   \Delta \hat{b}^{\text{CSS}}(\tau)^2 & = \frac{e^{2 \kappa(\tau)}+ 2 J-\sin(\theta)^2\,\sinh(2 \kappa(\tau))}{8\,J^3 T \tau\, \cos(\theta)^2\,\sin(\theta)^2} .\
\label{deltabgn}
\end{align}
The goal is to optimize Eq.\,(\ref{deltabgn}) with respect to encoding time and polar angle for fixed $2J$ and $T$. 

\smallskip

\paragraph{Markovian noise.}
In this limit, numerical minimization of Eq.\,(\ref{deltabgn}) with respect to $\{\theta, \tau\}$ shows that the optimal polar angle lies in the $\theta \ll 1$ limit. Taking the  $\theta$ derivative  of $\Delta \hat{b}(\tau)^2$ and equating to zero yields the condition
\begin{align*}
    2 J \sin(\gamma \tau)=e^{2 \gamma \tau } \cos(2 \theta) \sin(\theta)^{-4} \quad \Rightarrow \quad  2 J \sin(\gamma \tau) \approx e^{2 \gamma \tau } \theta^{-4} , \quad \theta \ll 1.
\end{align*}
Solving for the polar angle in this limit, we find $\theta_{\text{opt}}= (e^{2\gamma \tau}/(2 J \sinh(2 \gamma \tau))^{1/4}$. Replacing in Eq.\,(\ref{deltabgn}) and expanding to second order for $1/J \ll 1$ leads to
\begin{align}
    \Delta \hat{b}^{\text{CSS}}(\tau)^2 \approx \frac{\sinh(2 \gamma \tau)}{4 \tau T J^2} + \frac{\sqrt{e^{4 \gamma \tau}-1}}{4 \tau T J^{5/2}}. 
    \label{dbgn}
\end{align}
To determine the optimal encoding time, we now take the derivative of Eq.\,(\ref{dbgn}) with respect to $\tau$ and equate to zero, yielding \begin{align*}
    1-e^{4 \gamma \tau} (1-2 \gamma \tau)-\sinh(2 \gamma \tau)+ \sqrt{J (e^{4 \gamma \tau}-1) }=0 \quad \Rightarrow \quad -2 \gamma \tau + \frac{16}{3} \sqrt{ J \gamma \tau}\, (\gamma \tau)^3 \approx 0, 
\end{align*}
where the approximate equality comes from performing a Taylor series of the original equation in powers of $\gamma \tau \ll 1$ in the limit where $\sqrt{ J \gamma \tau}\gg 1$. We can then solve for the optimal time from this second expression. Replacing in $\theta_{\text{opt}}$, Eq.\,(\ref{dbgn}), and expanding to the leading powers of $J$ in the asymptotic limit, we arrive at
\begin{eqnarray}
   \Delta \hat{b}_{\text{opt}}^{\text{CSS}}= \frac{1}{\sqrt{2}}\, \sqrt{\frac{\gamma}{T}}\,J^{-1} ,\quad \tau_{\text{opt}}^{\text{CSS}}= \frac{1}{\gamma} \bigg( \frac{3^{2/5}}{2^{6/5}} J^{-1/5} - \frac{2^{2/5} 3^{1/5}}{5} J^{-3/5} \bigg), \quad 
   \theta^{\text{CSS}}_{\text{opt}}={2^{-1/5}\, 3^{-1/10}} \,J^{-1/5} .
   \label{dbCSS}
\end{eqnarray}
Comparison of these asymptotic expression with numerical optimization, as shown in Fig.\,\ref{ratio}, shows excellent agreement.
We further observe that the optimal uncertainty $\Delta \hat{b}_{\text{opt}}^{\text{CSS}}$ outperforms Eq.\,(\ref{dbCSSM}), where the polar angle remained fixed at the optimal noiseless value $\theta=\pi/4$, by a constant $1/\sqrt{2}$ factor. 

\smallskip

\paragraph{Zeno dynamics.}
In this case we may simply expand Eq.\,(\ref{deltabgn}) to second order in the encoding time, obtaining 
\begin{align}
    \Delta \hat{b}^{\text{CSS}}(\tau)^2 \approx \frac{{J}/{2}+2 J^2 \kappa_0^2 (\omega_c \tau)^2 \sin(\theta)^2}{T\, \tau\,J^4\,\sin(2\theta)^2} .
    \label{dbappth}
\end{align}
Equating the time derivative of Eq.\,(\ref{dbappth}) to zero and solving for the encoding time leads to the condition
\begin{align*}
    (\sin(\theta)^{-2}-4 J \kappa_0^2 (\omega_c \tau)^2 )=0 \quad \Rightarrow \quad \tau_{\text{opt}}=\frac{1}{2}\,|\sin(\theta)|^{-1}\,\frac{1}{ \kappa_0\,\omega_c} J^{-1/2}.
\end{align*}
Replacing into Eq.\,(\ref{dbappth}), we obtain
 $ \Delta \hat{b}(\tau_{\text{opt}})^2=2\,({ \kappa_0\omega_c}/{T})\,\sin(\theta)\,\sin(2 \theta)^{-2}\, J^{-5/2}$, which is easily minimized as a function of $\theta$. It follows that the optimal asymptotic uncertainty, encoding time, and polar angle are now given by:
\begin{align}
     \Delta \hat{b}_{\text{opt}}^{\text{CSS}}= \frac{3^{3/4}}{2}\, \sqrt{\frac{\omega_c}{T}}\,J^{-5/4 },
     \quad \tau_{\text{opt}}^{\text{CSS}}= \frac{\sqrt{3}}{2}\,\frac{1}{\kappa_0\omega_c}\, J^{-1/2}, \quad 
     \theta^{\text{CSS}}_{\text{opt}}=\text{arccot}(\sqrt{2}). 
     \label{dbCSSb}
\end{align}
Once again, the asymptotic expressions for $\tau_{\text{opt}}^{\text{CSS}}, \theta^{\text{CSS}}_{\text{opt}}$ are shown to correctly predict their $J \gg 1$ behavior see Fig.\,\ref{Ratio}. Note that $ \theta^{\text{CSS}}_{\text{opt}}$ is actually quite close to the optimal noiseless value, so $\Delta \hat{b}_{\text{opt}}^{\text{CSS}}$ is outperforming that of Eq.\,(\ref{zeno}) by a mere $0.95$ factor. Our results are in full agreement with the formulas in Table \ref{table:1} for a CSS, which were derived through a more general procedure.

\subsection{Ratio estimator performance}

Replacing the relevant mean values, Eqs.\,(\ref{Jxt})-\eqref{Jyt}, and Eqs.\,\eqref{JxSq2}-(\ref{JySq2}), into the squared uncertainty associated to the ratio estimator, Eq.\,(\ref{DeltabR}), by taking $ b_0=0, \phi=0$, we find
\begin{align}
    \Delta \hat{b}_{\text{R}}^{\text{CSS}}(\tau)^2=  \frac{e^{2 \kappa(\tau)} \cos(\theta)^2+\cosh(2 \kappa(\tau)) \sin(\theta)^2 + 2 J\sin(\theta)^2\,\sinh(2 \kappa(\tau))}{8\,J^3 T \tau\, \cos(\theta)^2\,\sin(\theta)^2}
    \label{dbR2}
\end{align}
Let us now optimize over both encoding time and polar angle to derive the best performance.

\smallskip

\paragraph{Markovian noise.} Numerical evidence shows that, in the $\theta \ll 1$ regime optimizing precision, the second term in the numerator of Eq.\,(\ref{dbR2}) may be safely discarded. The resulting expression,
\begin{align*}
  \Delta \hat{b}_{\text{R}}^{\text{CSS}}(\tau)^2 \approx    \frac{e^{2 \gamma \tau} \cos(\theta)^{2} + 2J\,\sinh(2 \gamma \tau)\,\sin(\theta)^{2}}{8\,T\,\tau\,J^3\,\cos(\theta)^2\,\sin(\theta)^2}, 
\end{align*}
is clearly upper-bounded by the uncertainty from measuring $J_y$, Eq.\,(\ref{deltabgn}) for $\kappa(\tau)=\gamma \tau$, to which it rapidly converges as $\cos(\theta)^2 \to 1$. It follows that the asymptotically optimal precision and parameters optimizing it are given by Eq.\,(\ref{dbCSS}), consistent with our numerical findings. 

\smallskip

\paragraph{Zeno dynamics.}
Expanding Eq.\,(\ref{dbR2}) up to second order in $\tau$ in the Zeno regime $\kappa(\tau) \approx \kappa_0^2 (\omega_c \tau)^2$, we find
\begin{align}
    \Delta \hat{b}_{\text{R}}^{\text{CSS}}(\tau)^2 & \approx \frac{1+2\,\kappa_0^2\,(\omega_c \tau)^2 (\cos(\theta)^2 + 2 J \sin(\theta)^2 )}{8\,T\,\tau\,J^3\,\cos(\theta)^2\,\sin(\theta)^2}  
    \label{dbRapp} .
\end{align}
Taking the time derivative of Eq.\,(\ref{dbRapp}) and equating to zero leads to the condition
\begin{align*}
 2 \kappa_0^2 (\omega_c \tau)^2 \big(\cos(\theta)^{2}+ 2J\,\sin(\theta)^2\big) - 1=0\quad \Rightarrow \quad 
 \tau_{\text{opt}}=\frac{1}{\sqrt{2 \kappa_0}\,\omega_c} 
 \left( \frac{\sin(\theta)^{2}+\cos(\theta)^{2}}{ \cos(\theta)^{2}+ 2J\,\sin(\theta)^2} \right)^{1/2}.
\end{align*}
Plugging $\tau_{\text{opt}}$ back in Eq.\,(\ref{dbRapp}) for $J \gg 1$, we find
\begin{align*}
    \Delta \hat{b}_R^{\text{CSS}}(\tau_{\text{opt}})^2 \approx \frac{2 \kappa_0 \omega_c}{T}\,\frac{\sin(\theta)}{\sin(2\theta)^2}\,J^{-5/2}, 
\end{align*}
which has a minimum for $\theta_{\text{opt}}= \text{arccot}(\sqrt{2})$. The optimal asymptotic performance for the ratio estimator, and the encoding time and polar angle realizing it are then given by:
\begin{align*}
      \Delta \hat{b}_{\text{R opt}}^{\text{CSS}}= \frac{3^{3/4}}{2}\, \sqrt{\frac{\omega_c}{T}}\,J^{-5/4} ,
      \quad \tau_{\text{R opt}}^{\text{CSS}}= \frac{\sqrt{3}}{2}\,\frac{1}{\kappa_0 \,\omega_c}\, J^{-1/2}, 
      \quad \theta^{\text{CSS}}_{\text{R opt}}=\text{arccot}(\sqrt{2}). 
\end{align*}
Comparing with Eq.\,(\ref{dbCSSb}), we conclude that the ratio estimator achieves the same precision scaling as the QCRB for non-Markovian, collective noise. This is confirmed by the numerical results presented in Fig.\,\ref{Ratio} in the main text.

\end{appendix}

\vspace*{10mm}

\twocolumngrid


\begin{thebibliography}{74}
\bibitem [{err()}]{erratum}%
  \BibitemOpen
  \href@noop {} {}\bibinfo {note} {Note that the method of moments uncertainty,
  Eq.\,(6) in Ref.\,\onlinecite{Bias}, has an additional incorrect $\nu^{-1}$
  factor.}\BibitemShut {Stop}%
\end{thebibliography}

\providecommand{\noopsort}[1]{}\providecommand{\singleletter}[1]{#1}%

\end{document}